%% file: ArXiv.tex
\documentclass[journal]{IEEEtran}

\usepackage{booktabs}

\usepackage{amssymb}
\usepackage{mathrsfs}

\usepackage{color}
\usepackage{cite}
\ifCLASSINFOpdf
  \usepackage[pdftex]{graphicx}
  \graphicspath{{../pdf/}{../jpeg/}}
   \DeclareGraphicsExtensions{.pdf,.jpeg,.png}
\else
  \usepackage[dvips]{graphicx}
\fi
\usepackage[cmex10]{amsmath}
\usepackage{algorithmic}
\usepackage{array}
\ifCLASSOPTIONcompsoc
  \usepackage[caption=false,font=normalsize,labelfont=sf,textfont=sf]{subfig}
\else
  \usepackage[caption=false,font=footnotesize]{subfig}
\fi
\usepackage{fixltx2e}
\usepackage[english]{algorithm2e}

\begin{document}
%\title{Inertia-Constrained Pixel-by-Pixel Nonnegative Matrix Factorisation: a Blind Source Separation Method Dealing with Source Variability}
\title{Inertia-Constrained Pixel-by-Pixel Nonnegative Matrix Factorisation: a Hyperspectral Unmixing Method Dealing with Intra-class Variability}
\author{Charlotte~Revel, %~\IEEEmembership{Member,~IEEE,}
        Yannick~Deville,~\IEEEmembership{Member,~IEEE,}
        V\'{e}ronique~Achard %~\IEEEmembership{Fellow,~OSA,}
        and~Xavier~Briottet %,~\IEEEmembership{Life~Fellow,~IEEE}% <-this % stops a space
\thanks{The research presented in this paper was funded by the Midi-Pyr\'{e}n\'{e}es region and the French ANR (ANR HYEP ANR 14-CE22-0016-01).}
\thanks{Ms. Revel and Mr. Deville are with the department of Signal, Image in Universe Sciences, IRAP, Toulouse University, 31400, France.}
\thanks{Ms. Revel, Ms. Achard and Mr. Briottet are with the department of Theoretical and Applied Optics, ONERA The French Aerospace Lab, Toulouse University, Toulouse, 31400, France.}}% <-this % stops a space
%\thanks{Manuscript received April 19, 2005; revised September 17, 2014.}}
\markboth{ArXiv}%
{Shell \MakeLowercase{\textit{et al.}}: Bare Demo of IEEEtran.cls for Journals}
\maketitle
\begin{abstract}
Blind source separation is a common processing tool to analyse the constitution of pixels of hyperspectral images. Such methods usually suppose that pure pixel spectra (endmembers) are the same in all the image for each class of materials. In the framework of remote sensing, such an assumption is no more valid in the presence of intra-class variabilities due to illumination conditions, weathering, slight variations of the pure materials, etc... In this paper, we first describe the results of investigations highlighting intra-class variability measured in real images. Considering these results, a new formulation of the linear mixing model is presented leading to two new methods. Unconstrained Pixel-by-pixel NMF (UP-NMF) is a new blind source separation method based on the assumption of a linear mixing model, which can deal with intra-class variability. To overcome UP-NMF limitations an extended method is proposed, named Inertia-constrained Pixel-by-pixel NMF (IP-NMF). For each sensed spectrum, these extended versions of NMF extract a corresponding set of source spectra. A constraint is set to limit the spreading of each source's estimates in IP-NMF. The methods are tested on a semi-synthetic data set built with spectra extracted from a real hyperspectral image and then numerically mixed. We thus demonstrate the interest of our methods for realistic source variabilities. Finally, IP-NMF is tested on a real data set and it is shown to yield better performance than state of the art methods.
\end{abstract}

% \fi
\IEEEpeerreviewmaketitle
\section{Introduction}
\label{Introduction}

\IEEEPARstart{H}{yperspectral} imaging is a common tool in the framework of remote sensing. Images provided by these sensors are spectrally highly resolved. However they lead to a decrease of the spatial resolution. Thus each signal recorded by a pixel is the result of a combination of spectra of several pure materials composing the area delimited by the pixel projected onto the ground. Retrieving these pure spectra, usually called endmembers, and each material proportion, called abundances, brings, in each considered pixel, interesting sub-pixel information. This source separation problem is called unmixing in remote sensing. How pure reflectance spectra are combined depends on the scene. The most common mixing model is the linear one \cite{keshava_spectral_2002}. The weight of each pure material spectrum in each sensed signal is proportional to the covered area. This model is adapted to flat macroscopic observed scenes \cite{singer_mars_1979} under a homogeneous irradiance. In case of 3D scene, nonlinear models are often more adapted, for instance a linear-quadratic mixing model was developed in \cite{meganem_linear_2014} for urban areas. For microscopic effects, intimate mixture models are used \cite{mustard_quantitative_1987}. Once the nature of the mixing model is known, source separation, also named unmixing in this study, can be performed. A large number of methods, requiring more or less prior knowledge, have been developed \cite{bioucas-dias_hyperspectral_2012}. However they hardly handle source variations. Source variations mean that the reflectance spectrum of a source of a same class can vary from an observation to another. For instance, the spectra of two similar roof materials in two different locations may differ. In remote sensing, this phenomenon is called intra-class variability. 
Various approaches \cite{zare_endmember_2014} have been developed to deal with this problem. Some of them make use of library \cite{roberts_hierarchical_2009}. Others described the intra-class variability as a statistical distribution and have proposed Bayesian methods, with various prior knowledges and models, to find the characteristics of this distribution and the associated abundances \cite{eches_bayesian_2010}, \cite{dobigeon_joint_2009}. However they frequently need particular knowledges. In a recent review of the unmixing methods dealing with intra-class variability \cite{zare_endmember_2014}, only two methods do not use prior knowledge or predefined source libraries. These two methods create a library from observed data and perform a sparse regression with this source library \cite{castrodad_learning_2011}. Also Somers et al., in \cite{somers_automated_2012}, proposed an unmixing method based on a library extraction. This type of approach is interesting since a library allows one to describe source variability. Indeed if the library is large enough it can contain several spectra of the same material. However it does not allow one to extract exact sources in each pixel signal. 

Here we propose a blind source separation method to extract from each observed signal the exact sources composing it. Our final method, called IP-NMF, is a constrained evolution of UP-NMF (Unconstrained Pixel-by-pixel NMF) that we first introduce. UP-NMF is based on an extended Nonnegative Matrix Factorisation (NMF) introduced by Lee and Seung in \cite{lee_learning_1999-1}. Section~\ref{Intra-class variability problems} evaluates the intra-class variability problems on a real image. Section \ref{Problem Statement} describes the general mixing model formulation ensuing from the observations developed in the previous part, Sections \ref{UP-NMF} and \ref{IP-NMF}, our extended UP-NMF and IP-NMF methods, Section \ref{Test Results on semi-synthetic data set}, their performance on a realistic data set and Section~\ref{Test Results on real images} the performance of IP-NMF on a real image.

\section{Intra-class variability}
\label{Intra-class variability problems}

\input{Intraclass_variability}

\section{Unmixing Problem Statement}
\label{Problem Statement}

\input{Problem_Statement}
\section{Unconstrained Pixel-by-pixel Nonnegative Matrix Factorisation (UP-NMF)}
\label{UP-NMF}

\input{UPNMF}

\section{Inertia-constrained Pixel-by-pixel Nonnegative Matrix Factorisation (IP-NMF)}
\label{IP-NMF}

\input{IPNMF}

\section{Test Results on semi-synthetic data set}
\label{Test Results on semi-synthetic data set}

\input{Test_results}

\section{Test Results for real image}
\label{Test Results on real images}

\input{Test_results_real_image}

\section{Conclusion}
In this paper, intra-class variability was first studied in order to characterise its magnitude. This analysis led us to develop a new mixing model taking this phenomenon into account. Two blind unmixing methods were then introduced to deal with intra-class variability. Tests on semi-synthetic data showed that UP-NMF has lower performance than IP-NMF. IP-NMF performance was also assessed with a real image. To process hyperspectral images, IP-NMF has the advantage of being automated, only the number of classes needs to be fixed. When initialisation of this method is provided by a usual unmixing method, IP-NMF always yields better performance than the latter method in our tests. Indeed IP-NMF brings its flexibility to find endmembers around the single spectrum per class used to initialise it. The sensitivity of IP-NMF to the initial number of classes is lower than that of the considered classical unmixing methods. Besides, IP-NMF is able to find classes which were not detected by the above classical methods. This makes IP-NMF more robust to processing real images. Besides, to increase the accuracy of the results and to exploit the large number of endmembers provided by IP-NMF a post-processing stage can be added. Moreover, if users of this method agree to act on the algorithm, a manual initialisation could also be used instead of the above automated initialisation. This kind of initialisation allows the users to choose their classes of interest.

Our method is an extension of NMF. We chose to use a gradient descent to obtain our estimated spectra and abundances. Variants of this method may be developed by using other optimisation algorithms. The cost function optimised in this version penalises the spread of classes. Other versions could be imagined which would penalise the cost function with other terms (distance to the initial spectrum, introduction of spatial constraints ...). We plan to develop such versions and to compare them with the UP-NMF and IP-NMF methods.   
%And a decrease of the initial spectra (in the context of unmixing) will affect the accuracy of this method. That is  We have developed a blind source separation method called IP-NMF to address source variability. IP-NMF is particularly adapted to the remote sensing blind inversion problem (also called blind unmixing). Tests on semi-synthetic data show improvements brought by this method and demonstrate the ability of IP-NMF to deal with realistic source variability. This investigation is a first step in the development of a more robust method. Indeed performance highly depends on initialisation. This point may be improved. Moreover the inertia constraint can be discussed to better take data spatial variability into account. Current work aims at developing an extended version of this approach both in terms of initialisation and constraint applied to the cost function. e

%\section*{Acknowledgements}
%The research presented in this paper is funded by ANR (ANR HYEP ANR 14-CE22-0016-01) and the Midi-Pyr\'{e}n\'{e}e region.

\ifCLASSOPTIONcaptionsoff
  \newpage
\fi

\bibliographystyle{IEEEtran}
\bibliography{Biblio.bib}

\end{document}

%% file: Intraclass_variability.tex
% Intra-class variability problems
%
	\subsection{Problem statement}
	\label{Problems}
	
\begin{figure}[h]
	\centering
	\includegraphics[scale=0.5]{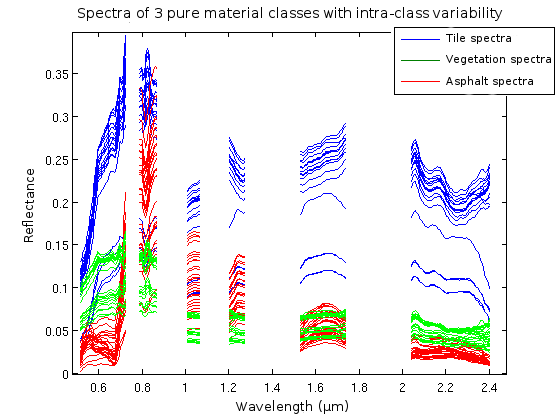}
	\caption{Spectra of 3 different classes of materials: vegetation (red), tile (blue), asphalt (green).}	
	\label{Spectres intra-class variability}
\end{figure}

The popular unmixing model described by Keshava and Mustard in \cite{keshava_spectral_2002} assumes that a scene can be fully reconstructed by using one spectrum per pure material present in the image, called an endmember. The notion of material is hard to define when the information of interest to be extracted from the scene has a macroscopic scale. Remote sensing frequently aims at studying landscapes, in this case what is called an endmember, or source, is a class of materials rather than a chemical material. For instance endmembers may be: grass, roads, tiles, trees... Now the grass is not the same from an area to another (various species, hydric stress...), tiles vary from one roof to another (various weatherings, mineral slight composition variations ...), and so on \cite{dennison_comparison_2004}, \cite{lacherade_spectral_2005}. So, at a macroscopic scale, an endmember is not the spectrum of a pure material but the spectrum which better characterises a class of materials. Spectra depicted in Figure~\ref{Spectres intra-class variability} corroborate this conclusion.

Using this new definition of an endmember, intra-class variability is the spectral variability between reflected spectra present in the image and the endmenber associated with them. Currently, intra-class variability is modelled in different ways. In \cite{veganzones_new_2014} and \cite{nascimento_vertex_2005} this variability is described by a scale factor. With this model the set $\mathcal{E}_{m}$ of spectra, present in the image, belonging to the $m^{th}$ source, is written as follows: 
$\mathcal{E}_{m} =  \lbrace \mathbf{s} \setminus \mathbf{s} = \gamma \mathbf{r_{m}}, \; \gamma \in \mathbb{R}^{\ast +} \rbrace$
where $\mathbf{r_{m}} \in \mathbb{R}^{L \times 1}$ is the $m^{th}$ endmember, $L$ the number of spectral bands. This model assumes that intra-class variability is only due to illumination variations for instance because of landscape slopes variation in non flat areas. However this model does not take into account other types of variability previously described: material can be weathered, composition variation, etc.... Variability can also be modelled as a statistical distribution when Bayesian approaches are used to perform the unmixing \cite{eches_bayesian_2010}. These models are highly parametric and require some strong assumptions. Other models describe variability as a bundle \cite{bateson_endmember_2000}. In the latter two cases the variability is not described by a factor, the models lead to a more complex representation of the variability.

An experimental characterisation of intra-class variability is needed to develop a mixing model and the associated unmixing method. In \cite{garcia-haro_new_2005}, a brief characterisation of this variability is made. However this study is limited to two classes of well-defined materials and aims at showing the benefits of spectra standardisation regarding the shadow processing.  
	
	\subsection{Data description}
	\label{Data description}
	
	Our spectral variability study is performed on an urban image. The studied area is located in the city of Toulouse, France. Toulouse city is composed of a large number of characteristic urban areas. Its downtown is typical from old cities in the South of France: tile roofs, low rise homes or big architectural monuments (cathedral, town hall...), very dense urbanisation, small streets or large avenues, green spaces (public gardens, squares...),... The suburbs are either residential (one-storey private houses or residential buildings) or industrial (large industrial buildings). In this study, we focus on images of the city center.  

The airborne campaign was carried out in October 2013 \cite{adeline_material_2013}. The hyperspectral instrument is composed of two cameras: the first one covering the Visible Near Infra-Red (VNIR) domain ($414$nm to $992$nm) with a $0.8$m ground sampling distance (GSD) and the second one in the Short Wave Infra-Red (SWIR) range ($980$nm to $2498$nm), with $1.8$m GSD. To work on similar GSD, the VNIR image is degraded to $1.8$m GSD and then registered with the SWIR image. It provides $405$ bands over the range $ \begin{bmatrix} 414, & 2498 \end{bmatrix}$ nm with a $1.8$m spatial resolution. An atmospheric compensation is applied to the data using the COCHISE tool \cite{miesch_direct_2005}. After applying COCHISE, spectra are reflectance ones. After removing the atmospheric absorption bands (water vapor, CO2,...), the studied image contains 214 spectral bands. Figure~\ref{Spectres intra-class variability} depicts some of these spectra.	

\begin{figure}[h]
	\centering
	\includegraphics[scale=0.30]{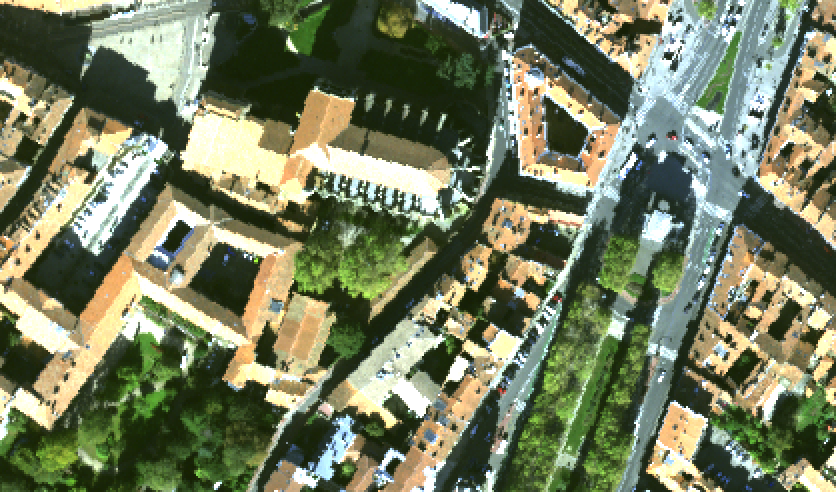}
	\caption{Sub-image selected for the intra-class variability study.}	
	\label{Image St Etienne}
\end{figure}

Spectra were extracted from a portion of the entire image. It was chosen by considering the presence of large homogeneous areas, i.e. areas entirely covered by a supposedly ``pure'' material. Figure~\ref{Image St Etienne} shows the selected sub-image. The roof of the cathedral, at the center of the image, is particularly interesting since it is covered by various tiles with various slopes.      
In this sub-image, various areas with similar materials were selected, considered as a class and then the corresponding spectra were manually collected by only keeping those which meet the following criteria:
\begin{itemize}
	\item[$\bullet$] pure material spectra
	\item[$\bullet$] various illumination conditions
	\item[$\bullet$] representative of the spectral variability of the materials due to composition, weathering (tiles of various roofs, asphalt extracted in different places...).
\end{itemize}
Three main classes of materials are considered: tiles, vegetation and asphalt. For these three classes, a large number of pure spectra can be extracted. It allowed relevant statistical studies. 

	\subsection{Data analysis}
	\label{Data study}

To characterise the intra-class variability, two measures were computed onto the extracted spectra: the correlation between spectra and the projection on the first Principal Components. 
Let $\mathbf{Y} = [\mathbf{y_{1}}, \ldots, \mathbf{y_{N}}]^{T}$ denote the matrix composed of a set of extracted spectra, with $\mathbf{y_{n}} \in \mathbb{R}^{L \times 1}$ a spectrum. The coefficients of the correlation matrix, $CORR_{i,j}$, can be written as follows:
\begin{equation*}
\label{Correlation}
CORR_{i,j} = \frac{Cov(\mathbf{y_{i}},\mathbf{y_{j}})}{\sigma_{y_{i}}\sigma_{y_{j}}}
\end{equation*} 
with $Cov(\cdot,\cdot)$ the covariance between two vectors and $\sigma_{y}$ the standard deviation of the vector $\mathbf{y}$.

\begin{figure}[h]
	\centering
	\includegraphics[scale=0.3]{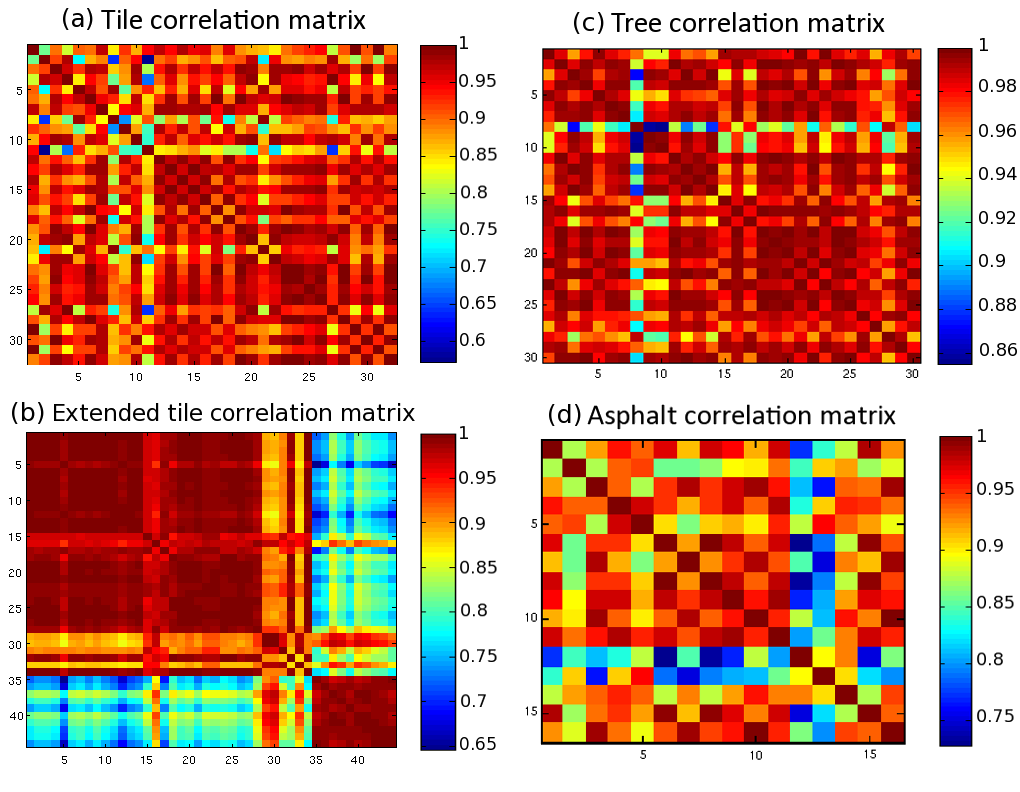}
	\caption{Pixel-to-pixel correlation matrices of the 3 classes (tile, tree and asphalt) showing the intra-class variability.}	
	\label{Corr classes}
\end{figure}

\begin{figure}[h]
	\centering
	\includegraphics[scale=0.20]{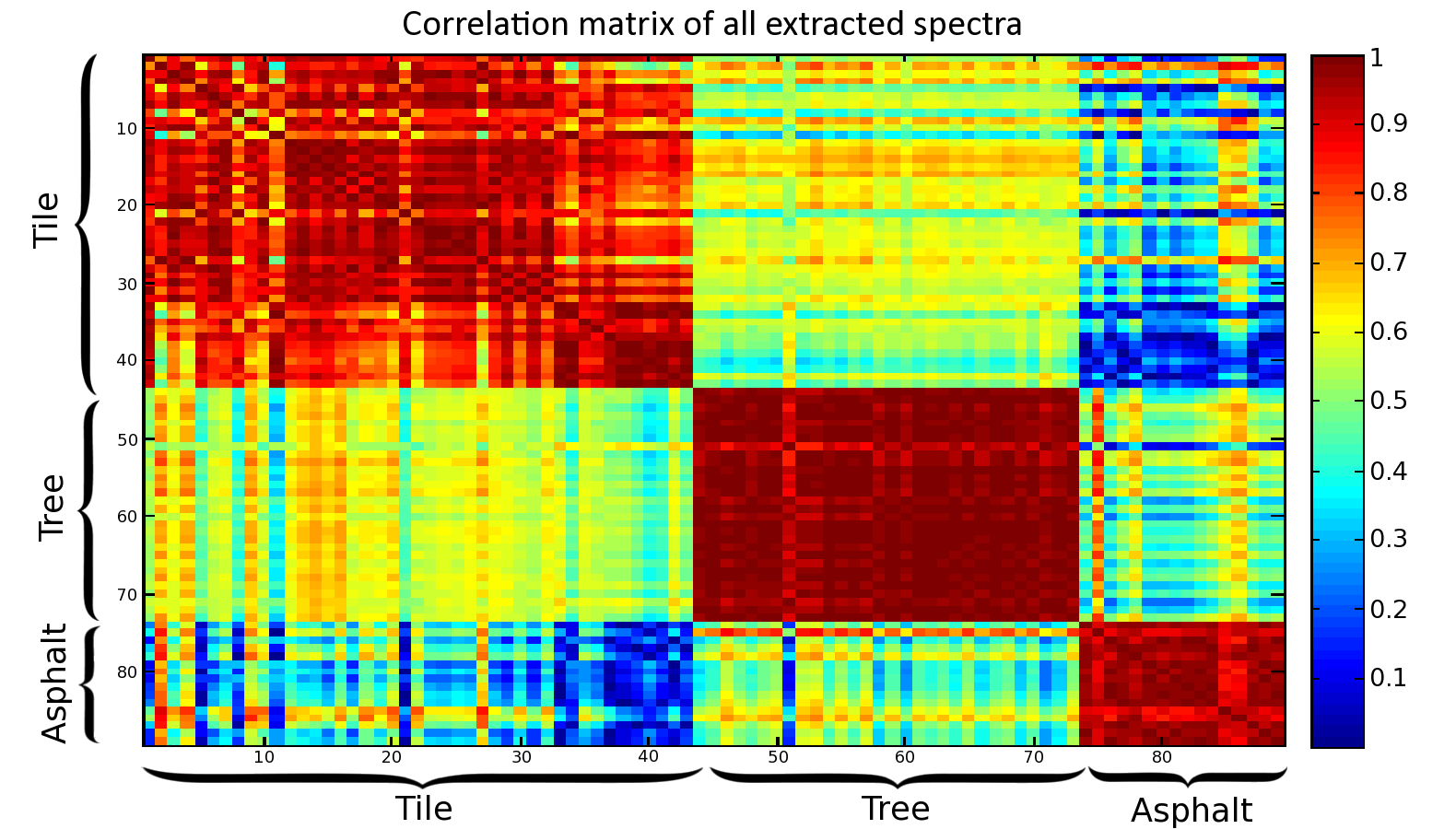}
	\caption{Correlation matrices of all extracted spectra.}	
	\label{Corr matrix}
\end{figure}

Figure~\ref{Corr classes} depicts four correlation matrices. For each of the tree and asphalt correlation matrices (Fig.~\ref{Corr classes} (c) and (d)), a single area of the sub-image corresponding to a single class was selected. The correlation was computed between all extracted spectra of this area. The tile correlation matrix (Fig.~\ref{Corr classes} (a)) was obtained in the same way, with tiles extracted from a same roof. The extended tile correlation matrix (Fig.~\ref{Corr classes} (b)) was computed between spectra extracted in various areas of the image, including the spectra used in the tile correlation matrix. \\%These matrices characterise the intra-class variability for 3 materials. \\
% For each one, an area of the sub-image corresponding to a single class was selected, the correlation was computed between all extracted spectra of this area.These matrices characterise the intra-class variability for 3 materials.
%
%The tile correlation factor matrix, in Figure~\ref{Corr classes}.(a), was computed from tile spectra extracted from the same roof. 
The correlation between the spectra extracted in a same roof fluctuates a lot, between $75\%$ and $95\%$, which shows the large variability of the class. The low correlation between some spectra (under $75\%$) is probably due to the non purity of some extracted spectra, resulting from various elements that can be fixed on roofs (gutter, aerial...). These extreme values were removed for the following studies.
%Since tiles were extracted from the same roof, we might think that the above low variability for tiles is due to the fact that intra-class variability is low in a small area (one building here) and becomes more considerable when we deal with larger areas (a whole image). 
Analyses of the asphalt correlation matrix are similar to the tile correlation matrix, the correlation between neighbouring asphalt spectra can be lower than $80\%$. The variability between spectra is due to different road surfacing. Results of the tree correlation matrices differ from the above results. The correlation between these spectra is higher than $90\%$ (which correspond to a low variability) except for the correlation of several spectra with the $8^{th}$ spectrum (probably a non pure spectrum). It can be due to the fact that in this area of Toulouse, most of the trees in streets belong to the same species, plane trees. The extended tile correlation matrix, in Fig.~\ref{Corr classes} (b), was obtained by computing pure spectra of a same material class extracted from various locations of the image. It appears that the correlation between spectra extracted from a same roof is higher, from $75\%$ to $95\%$, than the one between spectra of various roofs, from $65\%$ to $95\%$. The largest observed intra-class variability, in Fig.~\ref{Corr classes} and Fig.~\ref{Corr matrix}, makes sense as the tiles of Toulouse roofs are not the same (various mineral compositions...). It means that intra-class variability increases when the number of building, or independent areas, increases. This result can be extended to other classes of material like asphalt.      

Figure~\ref{Corr matrix} depicts the correlation matrix of all extracted spectra which illustrates both inter and intra-class variabilities.
Figure~\ref{Corr matrix} shows the variability at the sub-image scale, since spectra from the same material were extracted from various locations of the sub-image, as for the extended tile correlation matrix in Fig.~\ref{Corr classes}.  Firstly it appears, as depicted in Figure~\ref{Corr classes}, that the intra-class variability can be low (correlation higher than $80\%$ for the asphalt class). However it is higher than in Fig.~\ref{Corr classes} for some classes, especially when the class includes spectra extracted in various areas of the images, like tiles. The intra-class variability of a class depends on the observed scene. The tree class frequently has a high intra-class variability. It is not the case here (correlation around $90\%$) since the considered areas contain similar tree species.
The second important point is the high correlation between spectra of different classes. This variability is called inter-class variability. This phenomenon has to be considered. Indeed, if some classes have too close spectra, they would be hardly distinguishable especially when those classes have a high intra-class variability. For instance the correlation between some tile and vegetation spectra is higher than $60\%$ whereas some tile spectra are only $70\%$ correlated. This could lead to confusion between classes. Indeed if a class is defined by a bundle as in \cite{bateson_endmember_2000} the parameters have to be very carefully chosen to avoid spectra misclassification.

\begin{figure}[htp]
  %\hspace{-0.2cm}
  \centering
  \subfloat[Projection onto the $1^{st}$ and $2^{nd}$ axes]{\label{ACP_proj1}\includegraphics[scale=0.4]{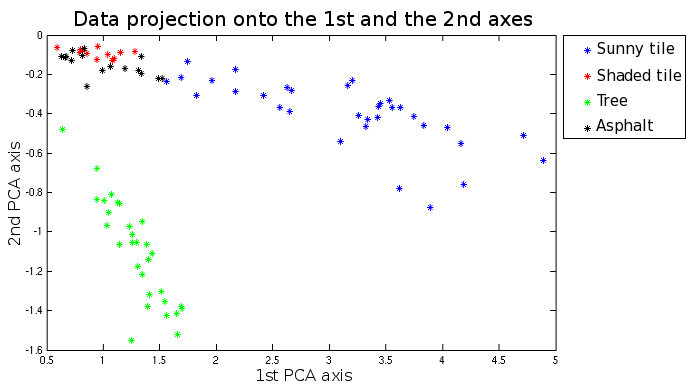}}
  
  \subfloat[Projection onto the $1^{st}$ and $3^{rd}$ axes]{\label{ACP_proj2}\includegraphics[scale=0.4]{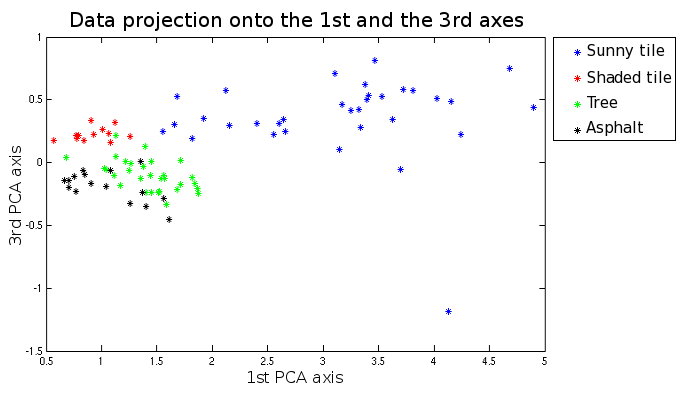}}
  \caption{Projection on the first PCA axes of 3 extracted spectra classes: tile (blue and red dots), tree (green dots), asphalt (black dots).}
  \label{ACP projection}
\end{figure}

The above correlation coefficients do not depict the variability of average level from one spectrum to another, that is highly related to illumination variation. To visualize it, a projection onto the first PCA axes was performed. Figure~\ref{ACP projection} illustrates the obtained results. In Figure~\ref{ACP projection}~(a) spectra of three classes are projected onto the first two principal axes. For the tile class, spectra are extracted from various roof slopes (i.e. various illumination conditions). The tree class is composed of spectra extracted from a same area, however the illumination of the canopy varies from a tree to another. This class also depicts the effects of illumination on the intra-class variability. The scale factor variability which cannot be seen in Figures~\ref{Corr classes} and~\ref{Corr matrix} appears clearly in Figure~\ref{ACP projection}. Indeed the set of spectra $\mathcal{E}_{m} =  \lbrace \mathbf{s} \setminus \mathbf{s} = \gamma \mathbf{r_{m}}, \; \gamma \in \mathbb{R}^{\ast +} \rbrace$ previously mentioned in subsection \ref{Problems} is represented in a PCA projection as a line passing through the origin. In Figure~\ref{ACP projection}~(b) this scale factor is visible for the sunny tile class. Spectra belonging to each class are more or less located along the same direction. However they do not exactly form a line. The variations on the axis passing through the origin correspond to the scale factor whereas the variations around this axis are linked to other phenomena. The location of blue and red dots in Figure~\ref{ACP projection} illustrates this. Indeed the darkest spectra (red dots) are the closest to the origin and the blue ones are situated further on the axis depending on their illumination. Asphalt projections corroborate this analysis. These spectra are dark ones and are located close to the origin and close to the spectra of dark trees and tiles.    

Taking these observations into account, it appears that the intra-class variability has to be considered to perform an accurate unmixing. These investigations also show that the modelling of the intra-class variability by a scale factor is not sufficient. So, in the following sections, a new mixing model dealing with intra-class variability is developed as well as unmixing methods adapted to this model.

%% file: Problem_Statement.tex
% Problem Statement
%
%We seek a technique capable of extracting sources in the case of a linear mixing model.
In the standard linear mixing model (LMM), each sensed spectrum $\mathbf{x_{p}} \in \mathbb{R}^{L \times 1}$ can be written as follows: 
\begin{equation}
\label{eq.LMM}
	\mathbf{x_{p}} = \sum_{m=1}^{M}c_{pm}\mathbf{r_{m}} \quad \forall p \in \lbrace 1, \ldots, P \rbrace 
\end{equation} 
where $p$ is the pixel index, $P$ the number of pixels, $m$ the index of one of the $M$ endmembers, $\mathbf{r_{m}} \in \mathbb{R}^{L \times 1}$ is the $m^{th}$ source spectrum and $c_{pm}$ is the associated mixing coefficient. The above model does not take into account intra-class variability. We introduce a new extended mixing model, which can be rewritten as:   
\begin{equation}
\label{eq.LMM var}
	\mathbf{x_{p}} = \sum_{m=1}^{M}c_{pm}\mathbf{r_{m}}(p) \quad \forall p \in \lbrace 1, \ldots, P \rbrace 
\end{equation} 
where $\mathbf{r_{m}}(p)$ is the spectrum associated with the $m^{th}$ source and the pixel $p$. Extracting these sources from such observations is an ill-posed problem. In Eq. (\ref{eq.LMM}) and (\ref{eq.LMM var}) we add the classical sum-to-one constraint (besides the nonnegativity constraint). This condition leads to:
\begin{equation}
\label{S1C}
	\sum_{m = 1}^{M}c_{pm} = 1 \quad \forall p \in \lbrace 1, \ldots, P \rbrace.
\end{equation} 
Let $\mathbf{X} = [\mathbf{x_{1}}, \ldots, \mathbf{x_{P}}]^{T}$ denote the sensed spectrum matrix, $\mathbf{R} = [\mathbf{r_{1}}, \ldots, \mathbf{r_{M}}]^{T}$ the endmember matrix if there is no intra-class variability (same sources for all pixels) and $\mathbf{C} = [\mathbf{c_{1}}, \ldots, \mathbf{c_{P}}]^{T}$ the mixing coefficient matrix. For all pixels $p$, $\mathbf{c_{p}} = [c_{p1}, \ldots, c_{pM}]^{T}$ is an $M$-element vector containing the set of mixing coefficients associated with the $p^{th}$ observed spectrum. The number of sources, $M$, is assumed to be known in the rest of the paper. The LMM can be written as follows:
\begin{equation}
\label{LMM matrix}
	\mathbf{X} = \mathbf{CR}.
\end{equation}   
To obtain a similar expression from model (\ref{eq.LMM var}), we introduce $\mathbf{R}(p)= [\mathbf{r_{1}}(p), \ldots, \mathbf{r_{M}}(p)]^{T}$, the set of $M$ sources associated with the observed spectrum $\mathbf{x}_{p}$, $\mathbf{\tilde{R}} = \begin{bmatrix} \mathbf{R}(1) \\ \ldots \\ \mathbf{R}(P) \end{bmatrix} \in \mathbb{R}^{PM \times L}$, the matrix containing all the sources and $\mathbf{\tilde{C}} \in \mathbb{R}^{P \times PM}$ a block-diagonal matrix denoting the new mixing coefficient matrix:
\begin{equation}
\label{C tilde}
	\mathbf{\tilde{C}} = \begin{bmatrix}
							\mathbf{c_{1}}^{T} & 0 \ldots 0 & \ldots & 0 \ldots 0 \\
							0 \ldots 0 & \mathbf{c_{2}}^{T} & \ldots & 0 \ldots 0 \\
							\; & \; & \ddots & \; \\
							0 \ldots 0 & 0 \ldots 0 & \ldots & \mathbf{c_{P}}^{T} \\	
	\end{bmatrix}.
\end{equation}
So, Eq. (\ref{eq.LMM var}) yields the matrix expression:
\begin{equation}
\label{LMM var matrix}
	\mathbf{X} = \mathbf{\tilde{C}\tilde{R}}.
\end{equation}
We aim at retrieving matrices $\mathbf{\tilde{C}}$ and $\mathbf{\tilde{R}}$ under the nonnegativity and sum-to-one constraints. 

%% file: UPNMF.tex
% UPNMF
%
%
%
%However this method does not take into account the desired closeness of estimated source from the same class in all sensed spectra. To overcome this limitation and the strong model underdetermination, we then add an inertia constraint to the cost function in Section~\ref{IP-NMF}. This means that the spreading of estimated sources from the same class is penalized.

Nonnegative Matrix Factorisation (NMF) has been adapted to solve remote sensing unmixing problems corresponding to the LMM \cite{qian_hyperspectral_2011-1}, \cite{hoyer_non-negative_2004}. NMF aims at decomposing a matrix in a product of two nonnegative matrices, the coefficient matrix, $\mathbf{C}$ and the reflectance matrix, $\mathbf{R}$. The necessary assumption of this method is the nonnegativity of the two searched matrices. The sum-to-one constraint can be added to NMF. The method can, then, be applied to the observation matrix.
To perform NMF a cost function, hereafter called $J_{nmf}$, is minimised, however the obtained minima may be local ones. This point can be overcome with a good initialisation. This will be discussed in a later part (Sec.~\ref{Test Results on semi-synthetic data set}).

Now considering the extended mixing model (\ref{LMM var matrix}), we introduce two evolutions of NMF for estimating $\mathbf{\tilde{C}}$ and $\mathbf{\tilde{R}}$. These two methods are respectively defined in this section and in Section~\ref{IP-NMF}.         
 
	\subsection{Cost function}
	\label{Cost function UP-NMF}	
	
Standard NMF aims at minimising the reconstruction error (RE) to extract a set of endmembers from a set of observed spectra. The standard cost function can be written as: 
\begin{equation}
\label{J_NMF}
J_{nmf} = \frac{1}{2} \begin{Vmatrix} \mathbf{X} - \mathbf{C}\mathbf{R} \end{Vmatrix}_{F}^{2}
\end{equation} 
where $\begin{Vmatrix} . \end{Vmatrix}_{F}$ is the Frobenius norm. In (\ref{J_NMF}), $\mathbf{C}$ and $\mathbf{R}$ represent the \emph{adaptive} variables which respectively aim at estimating the \emph{actual} coefficients and source spectra involved in the mixing model (\ref{LMM matrix}). By using $\mathbf{\tilde{C}}$ and $\mathbf{\tilde{R}}$ instead of $\mathbf{C}$ and $\mathbf{R}$ we extract one set of sources from each sensed spectrum. The Unconstrained pixel-by-pixel NMF (UP-NMF) method is an evolution of standard NMF which minimises this new cost function:
\begin{equation}
\label{J_UPNMF}
J_{upnmf} = \frac{1}{2} \begin{Vmatrix} \mathbf{X} - \mathbf{\tilde{C}}\mathbf{\tilde{R}} \end{Vmatrix}_{F}^{2}.
\end{equation}
As in (\ref{J_NMF}), $\mathbf{\tilde{C}}$ and $\mathbf{\tilde{R}}$ are \emph{adaptive} variables which aim at estimating the \emph{actual} coefficients and source spectra in the mixing model (\ref{LMM var matrix}). 
	\subsection{Gradient calculation}
	\label{Gradient calculation}
To minimise $J_{upnmf}$, we developed an extended version of Lin's standard NMF algorithm \cite{lin_projected_2007}. Lin's method is based on projected gradient. So, we have to determine the derivatives of $J_{upnmf}$ with respect to $\mathbf{\tilde{R}}$ and $\mathbf{\tilde{C}}$. These derivatives are easily calculated with the matrix formulas in \cite{petersen_matrix_2012}.
\begin{equation*}
	J_{upnmf} = \frac{1}{2}Tr((\mathbf{X} - \mathbf{\tilde{C}}\mathbf{\tilde{R}})(\mathbf{X} - \mathbf{\tilde{C}}\mathbf{\tilde{R}})^{T})
\end{equation*}
\begin{align}
\label{deriv.J1-R}
	\Rightarrow \frac{\partial J_{upnmf}}{\partial \mathbf{\tilde{R}}} & = -\mathbf{\tilde{C}}^{T}(\mathbf{X} - \mathbf{\tilde{C}}\mathbf{\tilde{R}}) \\
\label{deriv.J1-C}
	\frac{\partial J_{upnmf}}{\partial \mathbf{\tilde{C}}} & = -(\mathbf{X} - \mathbf{\tilde{C}}\mathbf{\tilde{R}})\mathbf{\tilde{R}}^{T}.
\end{align}
	\subsection{Update algorithm}
	\label{Update algorithm UP-NMF}
The above calculations allow us to formulate the UP-NMF update algorithm. We choose to use a gradient descent algorithm. The update rules for $\mathbf{\tilde{R}}$ and $\mathbf{\tilde{C}}$ are the following:
\begin{align}
\label{update rule R}
	\mathbf{\tilde{R}}^{(i+1)} & \longleftarrow \mathbf{\tilde{R}}^{(i)} - \alpha_{\mathbf{\tilde{R}}}\frac{\partial J_{upnmf}^{(i)}}{\partial \mathbf{\tilde{R}}} \\ 
\label{update rule C}
	\mathbf{\tilde{C}}^{(i+1)} & \longleftarrow \mathbf{\tilde{C}}^{(i)} - \alpha_{\mathbf{\tilde{C}}}\frac{\partial J_{upnmf}^{(i)}}{\partial \mathbf{\tilde{C}}}
\end{align} 
followed by projection onto $\mathbb{R}^{+*}$ and sum-to-one normalisation. $\alpha_{\mathbf{\tilde{R}}}$ and $\alpha_{\mathbf{\tilde{C}}}$ are positive adaptation steps. \\

Using (\ref{deriv.J1-R}) and (\ref{deriv.J1-C}), we obtain the update rules for our UP-NMF method. However only all $\mathbf{c_{p}}^{T}$ in (\ref{C tilde}) should thus be updated, whereas the other terms of $\mathbf{\tilde{C}}$ are kept to zero. So, the complete UP-NMF update can be written as follows: \\  
%
%\vspace{0.5cm}
%
\begin{algorithm}
$\quad$ 1. Update of matrix $\mathbf{\tilde{R}}$\\
$\mathbf{\tilde{R}}^{(i+1)} \longleftarrow \mathbf{\tilde{R}}^{(i)} + \alpha_{\tilde{R}}(\mathbf{\tilde{C}}^{(i)T}(\mathbf{X}-\mathbf{\tilde{C}}^{(i)}\mathbf{\tilde{R}}^{(i)}))$ \\
$\mathbf{\tilde{R}}^{(i+1)} = max(\mathbf{\tilde{R}}^{(i+1)}, \epsilon)$ \\
$\;$
$\quad$ 2. Update of matrix $\mathbf{\tilde{C}}$ \\
\For{$p = 1$ to $P$}
%\For{$p = 1$ to $P$} 
	{$\mathbf{c_{p}}^{(i+1)T} \longleftarrow \mathbf{c_{p}}^{(i)T}$ \\
$\quad \quad \quad \quad \quad \quad \quad +  \alpha_{\mathbf{\tilde{C}}}(\mathbf{x_{p}}^{T}-\mathbf{c_{p}}^{(i)T}\mathbf{R}(p)^{(i)})\mathbf{R}(p)^{(i)T}$ \\
	 $\mathbf{c_{p}}^{(i+1)} = max(\mathbf{c_{p}}^{(i+1)}, \epsilon)$}
%\EndFor
$\;$
$\quad$ 3. Normalisation of the coefficients \\
\For{$p = 1$ to $P$}
	{$\mathbf{c_{p}}^{(i+1)} \longleftarrow \mathbf{c_{p}}^{(i+1)} / \sum_{m=1}^{M}c_{pm}$}
%\EndFor
where $\epsilon$ is a small positive constant.
\end{algorithm}

%\vspace{1.5cm}

Due to the high underdeterminacy of this optimisation problem, the behavior of UP-NMF is not accurate enough. Estimated spectra $\mathbf{r_{m}}(p)$ from a same class $m$ may evolve so differently that they tend to define several classes of materials. This observation is all the more important that the inter-class variability is small, as it was explained in Sec.~\ref{Data study}. To limit this spreading of estimated spectra from a same class, constraints are required in the cost function. Such a constraint is proposed hereafter, Sec.~\ref{IP-NMF}. It is based on the observations made in Sec.~\ref{Data study} and aims at penalising the spreading of estimated sources from the same class.

%% file: IPNMF.tex
% IPNMF
%
	\subsection{Cost function}

Our second method is based on limiting class inertia to reduce the risk for an estimated pure spectrum to go out of its own class. This limitation is introduced in the optimisation problem by adding a penalty term in the cost function. The function $J_{upnmf}$ to be minimised thus becomes:
\begin{equation}
\label{J}
	J_{ipnmf} = \frac{1}{2} \begin{Vmatrix} \mathbf{X} - \mathbf{\tilde{C}}\mathbf{\tilde{R}} \end{Vmatrix}_{F}^{2} + \mu \sum_{m = 1}^{M} Tr(Cov(\mathbf{\tilde{R}_{\mathcal{C}_{m}}}))
\end{equation}
where $\mathbf{\tilde{R}_{\mathcal{C}_{m}}} \in \mathbb{R}^{P \times L}$ is the matrix containing estimates of all the endmembers of the $m^{th}$ class (i.e. spectra of the $m^{th}$ material), $Tr(Cov(\mathbf{\tilde{R}_{\mathcal{C}_{m}}}))$ is the inertia of the $m^{th}$ class and $\mu$ the constraint parameter. We compute $\sum_{m=1}^{M}Tr(Cov(\mathbf{\tilde{R}_{\mathcal{C}_{m}}}))$ rather than $Tr(Cov(\mathbf{\tilde{R}}))$ because the latter expression tends to aggregate all classes. 
The covariance matrix trace measures the spreading of the spectra constituting the matrix, whatever the gravity center is. By using this penalty term, spectra moving too far away from others penalise all the class. Thus homogeneity is preserved. So it can be expected that the resulting spectra which evolved from the same ``seed'' still define the same class.

	\subsection{Gradient calculation}
To minimise $J_{ipnmf}$, we develop an extended version of Lin's standard NMF algorithm \cite{lin_projected_2007}, as for UPNMF. So, here again, we have to determine the derivatives of the cost function with respect to $\mathbf{\tilde{R}}$ and $\mathbf{\tilde{C}}$. We start by decomposing $J_{ipnmf}$ in two terms:
\begin{eqnarray}
\label{JJ}
J_{ipnmf} & = & J_{RE} + \mu J_{I}
\end{eqnarray}
with
%
% \begin{eqnarray*}
\begin{align}
J_{RE} & = \frac{1}{2}\begin{Vmatrix} \mathbf{X} - \mathbf{\tilde{C}}\mathbf{\tilde{R}} \end{Vmatrix}_{F}^{2} \\
J_{I} & = \sum_{m = 1}^{M} Tr(Cov(\mathbf{\tilde{R}_{\mathcal{C}_{m}}})) \notag \\
\label{X}
\; & = \sum_{m = 1}^{M} (\frac{1}{P}Tr(\mathbf{\tilde{R}_{\mathcal{C}_{m}}}^{T}\mathbf{\tilde{R}_{\mathcal{C}_{m}}}) - \frac{1}{P^{2}}Tr(Q_{\mathcal{C}_{m}})) \\
Q_{\mathcal{C}_{m}} & = \mathbf{\tilde{R}_{\mathcal{C}_{m}}}^{T} \mathbf{1}_{P,P} \mathbf{\tilde{R}_{\mathcal{C}_{m}}}
\end{align}
%\end{eqnarray*}
%
\input{Calculs_IPNMF}

	\subsection{Update algorithm}

\input{Update_algo}

%% file: Calculs_IPNMF.tex
%% Calcul derivee de J
%
where $\mathbf{1}_{P,P}$ is a $P\times P$ dimensional matrix of ones. $J_{RE}$ defines the reconstruction error (RE) and $J_{I}$ the Inertia constraint. It can be noted that $J_{RE}$ is equal to $J_{upnmf}$, so the calculations for this term have already been made.
$J_{I}$ does not depend on $\mathbf{\tilde{C}}$ so the corresponding derivative is zero. To obtain the derivative of $J_{I}$ with respect to $\mathbf{\tilde{R}}$ we have to make use of scalar writing. To this end, we start by rewriting $J_{I}$ in order to extract $\mathbf{\tilde{R}}$. Eq. (\ref{X}) yields:
\begin{align}
J_{I} & = \sum_{m = 1}^{M} (\frac{1}{P}\sum_{l = 1}^{L}\sum_{k = 1}^{P}[\tilde{R}_{\mathcal{C}_{m}}]_{k,l}^{2} - \frac{1}{P^{2}}Tr(Q_{\mathcal{C}_{m}})) \notag \\
\; & = \frac{1}{P}\sum_{m = 1}^{M}\sum_{l = 1}^{L}\sum_{k = 1}^{P}[\tilde{R}]_{(k-1)M+m,l}^{2} - \frac{1}{P^{2}}\sum_{m = 1}^{M}Tr(Q_{\mathcal{C}_{m}}) \notag \\
\; & = \frac{1}{P}\sum_{\kappa = 1}^{PM}\sum_{l = 1}^{L}[\tilde{R}]_{\kappa,l}^{2} - \frac{1}{P^{2}}\sum_{m = 1}^{M}Tr(Q_{\mathcal{C}_{m}}) \notag \\
\label{J2_1}
\; & = \frac{1}{P}Tr(\mathbf{\tilde{R}}^{T}\mathbf{\tilde{R}}) - \frac{1}{P^{2}}\sum_{m = 1}^{M}Tr(Q_{\mathcal{C}_{m}}).
\end{align}
We note that the first term can be derived by using the matrix formulas in \cite{petersen_matrix_2012}:
\begin{equation}
\label{deriv.1}
	\frac{\partial}{\partial \mathbf{\tilde{R}}}(\frac{1}{P}Tr(\mathbf{\tilde{R}}^{T}\mathbf{\tilde{R}})) = \frac{2}{P}\mathbf{\tilde{R}}.
\end{equation}
Now focusing on the second term of Eq. (\ref{J2_1}), we introduce: \\
\begin{align}
\mathbf{A} & = \mathbf{\tilde{R}}_{\mathcal{C}_{m}}^{T} \longrightarrow a_{ij} = [\tilde{R}_{\mathcal{C}_{m}}]_{ji} \notag \\
\mathbf{B} & = \mathbf{1}_{P,P} \mathbf{\tilde{R}}_{\mathcal{C}_{m}} \longrightarrow b_{ij} = \sum_{\beta = 1}^{P}[\tilde{R}_{\mathcal{C}_{m}}]_{\beta j}. \notag
\end{align}
Then $Q_{\mathcal{C}_{m}} = \mathbf{A} \mathbf{B}$ and \\
\begin{align}
[Q_{\mathcal{C}_{m}}]_{ij} & = \sum_{\alpha = 1}^{P}a_{i\alpha}b_{\alpha j} \notag \\  
 & = \sum_{\alpha = 1}^{P}\sum_{\beta = 1}^{P}[\tilde{R}_{\mathcal{C}_{m}}]_{\alpha i}[\tilde{R}_{\mathcal{C}_{m}}]_{\beta j}. \notag
\end{align}
Therefore:
\begin{align}
	\label{Tr(Cm)} 
Tr(Q_{\mathcal{C}_{m}}) & = \sum_{l = 1}^{L}[Q_{\mathcal{C}_{m}}]_{l,l} \notag\\ 
 & = \sum_{l = 1}^{L}\sum_{\alpha = 1}^{P}\sum_{\beta = 1}^{P}[\tilde{R}_{\mathcal{C}_{m}}]_{\alpha \, l}[\tilde{R}_{\mathcal{C}_{m}}]_{\beta \, l}. 
\end{align}
From Eq. (\ref{Tr(Cm)}) we can calculate the derivative of $\sum_{m = 1}^{M} Tr(Q_{\mathcal{C}_{m}})$ with respect to $[\tilde{R}]_{\gamma \lambda}$, one element of $\mathbf{\tilde{R}}$. By definition of $\mathbf{\tilde{R}}_{\mathcal{C}_{m}}$, $[\tilde{R}]_{\gamma \lambda}$ is present in only one of the matrices $Q_{\mathcal{C}_{m}}$, i.e. the one with $m = 1 +(\gamma - 1)(mod \; M)$, denoted as $\eta$ hereafter.%We introduce $\mu \equiv \gamma[M]$ to define this $m$. 
\begin{align}
\label{deriv.2}
	\frac{\partial}{\partial [\tilde{R}]_{\gamma \lambda}}(\sum_{m = 1}^{M} Tr(Q_{\mathcal{C}_{m}})) & = \sum_{m = 1}^{M}\frac{\partial}{\partial [\tilde{R}]_{\gamma \lambda}}(Tr(Q_{\mathcal{C}_{m}})) \notag \\
 	& = \frac{\partial}{\partial [\tilde{R}]_{\gamma \lambda}}(Tr(Q_{\mathcal{C}_{\eta}})) \notag \\
 	& = \frac{\partial}{\partial [\tilde{R}]_{\gamma \lambda}}(\sum_{l = 1}^{L}\sum_{\alpha = 1}^{P}\sum_{\beta = 1}^{P}[\tilde{R}_{\mathcal{C}_{\eta}}]_{\alpha \, l}[\tilde{R}_{\mathcal{C}_{\eta}}]_{\beta \, l}). 
\end{align}
From Eq. (\ref{deriv.2}), four cases are possible:
\begin{equation*}
\frac{\partial (\sum_{l = 1}^{L} [\tilde{R}_{\mathcal{C}_{\eta}}]_{\alpha \, l}[\tilde{R}_{\mathcal{C}_{\eta}}]_{\beta \, l})}{\partial [\tilde{R}]_{\gamma \lambda}} = \begin{cases}
	0 & \text{if $\alpha \neq \gamma$ and $\beta \neq \gamma$}, \\
	[\tilde{R}_{\mathcal{C}_{\eta}}]_{\beta \, \lambda} & \text{if $\alpha = \gamma$ and $\beta \neq \gamma$}, \\
	[\tilde{R}_{\mathcal{C}_{\eta}}]_{\alpha \, \lambda} & \text{if $\alpha \neq \gamma$ and $\beta = \gamma$}, \\
	2[\tilde{R}_{\mathcal{C}_{\eta}}]_{\alpha \, \lambda} & \text{if $\alpha = \beta = \gamma$}, 
\end{cases}
\end{equation*}
therefore:
\begin{align}
\label{deriv.3}
	\frac{\partial}{\partial [\tilde{R}]_{\gamma \lambda}}(\sum_{m = 1}^{M} Tr(Q_{\mathcal{C}_{m}})) & = 2\sum_{\alpha = 1}^{P}[\tilde{R}_{\mathcal{C}_{\eta}}]_{\alpha \, \lambda} \notag  \\
	& = 2 \sum_{\alpha = 1}^{P}[\tilde{R}]_{(\alpha -1)M + \eta , \lambda} .
\end{align}
Result (\ref{deriv.3}) can be extended to all entries of $\mathbf{\tilde{R}}$, which yields:
\begin{equation}
	\label{deriv.4}
	\frac{\partial}{\partial \mathbf{\tilde{R}}}(\sum_{m = 1}^{M} Tr(Q_{\mathcal{C}_{m}})) = 2 \mathbf{U \tilde{R}}
\end{equation}
with $\mathbf{U} \in \mathbb{R}^{PM \times PM}$ defined by 
\begin{eqnarray*}
	\mathbf{U} = 
	\begin{bmatrix}
		 \begin{matrix} 
		 	M \left\{
		 	\begin{matrix}
		 		\, \\
		 		\, \\
		 		\, \\
		 		\,
		 	\end{matrix} \right. \\
		 	\, \\
		 	\,
		 \end{matrix}
		  \overbrace{
			\begin{matrix}
				1 & 0 & \ldots & 0 \\
				0 & 1 & \ldots & 0 \\
				\vdots & \; & \ddots & \; \\
				0 & 0 & \ldots & 1 \\ 
				1 & 0 & \ldots & 0 \\
				\vdots & \; & \; & \;		
			\end{matrix}
		}^{M}  &
		\begin{matrix}
			1 & \ldots \\
			0 & \ldots \\
			\vdots & \; \\
			0 & \ldots \\
			1 & \ldots \\
			\; & \ddots
		\end{matrix} 	
	\end{bmatrix}	
	=   
	\begin{bmatrix}
		\mathbf{Id}_{M} & \ldots & \mathbf{Id}_{M} \\
		\vdots & \ddots & \vdots \\
		\mathbf{Id}_{M} & \ldots & \mathbf{Id}_{M} 
	\end{bmatrix}.
\end{eqnarray*}
The notation $\mathbf{Id}_{D}$ stands for the $D$-dimensional identity matrix. Thanks to (\ref{J2_1}), (\ref{deriv.1}) and (\ref{deriv.4}) we obtain the partial derivatives of $J_{I}$:
\begin{equation}
\label{deriv.J2}
	\frac{\partial J_{I}}{\partial \mathbf{\tilde{R}}} = \frac{2}{P}(\mathbf{Id}_{PM} - \frac{1}{P}\mathbf{U})\mathbf{\tilde{R}}.
\end{equation}
By combining (\ref{deriv.J1-R}), (\ref{deriv.J1-C}) and (\ref{deriv.J2}) we obtain the two partial derivatives of the general cost function (\ref{JJ}) with respect to $\mathbf{\tilde{R}}$ and $\mathbf{\tilde{C}}$:
\begin{align}
	\frac{\partial J_{ipnmf}}{\partial \mathbf{\tilde{R}}} & = -\mathbf{\tilde{C}}^{T}(\mathbf{X} - \mathbf{\tilde{C}}\mathbf{\tilde{R}}) + \frac{2\mu}{P}(\mathbf{Id}_{PM} - \frac{1}{P}\mathbf{U})\mathbf{\tilde{R}} \notag \\
	\frac{\partial J_{ipnmf}}{\partial \mathbf{\tilde{C}}} & = -(\mathbf{X} - \mathbf{\tilde{C}}\mathbf{\tilde{R}})\mathbf{\tilde{R}}^{T}. \notag
\end{align}

%% file: Update_algo.tex
% Ecriture de la mise à jour des algorithmes 
%
The previous calculations allow us to formulate the update algorithm. The general gradient descent formulation was already given in (\ref{update rule R}) and (\ref{update rule C}). Using it we obtain the update rules for our IP-NMF method. As for UP-NMF, only the $\mathbf{c_{p}}^{T}$ in (\ref{C tilde}) should thus be updated, whereas the other terms of $\mathbf{\tilde{C}}$ are kept to zero. So, the complete IP-NMF update can be written as follows: \\  
\begin{algorithm}
\vspace{-0.2cm}
$\quad$ 1. Update of matrix $\mathbf{\tilde{R}}$\\
$\mathbf{\tilde{R}}^{(i+1)} \longleftarrow \mathbf{\tilde{R}}^{(i)} + \alpha_{\tilde{R}}(\mathbf{\tilde{C}}^{(i)T}(\mathbf{X}-\mathbf{\tilde{C}}^{(i)}\mathbf{\tilde{R}}^{(i)})$ \\
$ \quad \quad \quad \quad \quad \quad \quad \quad \quad \quad \quad \quad- \frac{2\mu}{P}(\mathbf{Id}_{PM} - \frac{1}{P}\mathbf{U})\mathbf{\tilde{R}}^{(i)})$\\
$\mathbf{\tilde{R}}^{(i+1)} = max(\mathbf{\tilde{R}}^{(i+1)}, \epsilon)$ \\
$\;$
\end{algorithm}
\begin{algorithm}
$\quad$ 2. Update of matrix $\mathbf{\tilde{C}}$ \\
\For{$p = 1$ to $P$}
%\For{$p = 1$ to $P$} 
	{$\mathbf{c_{p}}^{(i+1)T} \longleftarrow \mathbf{c_{p}}^{(i)T}$ \\
$\quad \quad \quad \quad \quad \quad \quad + \alpha_{\tilde{C}}(\mathbf{x_{p}}^{T}-\mathbf{c_{p}}^{(i)T}\mathbf{R}(p)^{(i)})\mathbf{R}(p)^{(i)T}$ \\
	 $\mathbf{c_{p}}^{(i+1)} = max(\mathbf{c_{p}}^{(i+1)}, \epsilon)$}
%\EndFor
$\;$
$\quad$ 3. Normalisation of the coefficients \\
\For{$p = 1$ to $P$}
	{$\mathbf{c_{p}}^{(i+1)} \longleftarrow \mathbf{c_{p}}^{(i+1)} / \sum_{m=1}^{M}c_{pm}$}
%\EndFor
\end{algorithm}

As all iterative algorithms, IP-NMF as UP-NMF, have to be initialised. The choices we made are described in the following Section~\ref{Test description SS}.

%% file: Test_results.tex
% Test_results
%
	\subsection{Test description}
	\label{Test description SS}

Tests of the UP-NMF and IP-NMF methods are firstly performed with semi-synthetic data. The sources are real spectra extracted from the hyperspectral sub-image described in Sec.~\ref{Data description}, they are depicted in Figure.~\ref{Spectres intra-class variability}. They describe realistic source variations. Mixing coefficients are randomly chosen while respecting the sum-to-one constraint. The mixing model is defined by Eq.~(\ref{eq.LMM var}). 

Tests are performed by varying the initialisations $\mathbf{\tilde{R}}^{(0)}$ and $\mathbf{\tilde{C}}^{(0)}$ and the constraint parameter $\mu$. Three $\mathbf{\tilde{R}}^{(0)}$ are tested: (i) $M$ mixed signals randomly selected from the observations, (ii) the $M$ purest signals extracted with a classical remote sensing blind source separation method (N-FINDR \cite{winter_n-findr:_1999}), (iii)~for each class $m$, the average of all $P$ source signals in this class. Mixing coefficient initialisation is obtained in two ways: (a) by giving the same constant value, $\frac{1}{M}$, to all coefficients, (b) by extracting coefficients associated with initialised spectra with a Fully Constrained Least Square (FCLS) regression \cite{heinz_fully_2001}, \cite{zhang_hybrid_2010}, \cite{broadwater_hybrid_2007}. The constraint parameter $\mu$ varies from $0$ to $100$ to assess the impact of $\mu$ on the algorithm performance. 

Among the above listed initialisations, only the results for one couple are discussed in the first part of Section \ref{Results SS}, namely $\mathbf{\tilde{R}}^{(0)}$ and $\mathbf{\tilde{C}}^{(0)}$ respectively initialised by (ii) and (a). Indeed (i) is a more uncertain initialisation than (ii) and (iii) is the best expected initialisation of $\mathbf{\tilde{R}}^{(0)}$ but it requires knowledge about the data which is not available for real data sets. (a) was chosen to initialise $\mathbf{\tilde{C}}^{(0)}$ to reduce the risk of being in a local minimum (this conclusion results from a study of UP-NMF and IP-NMF initialised with (b)).

	\subsection{Evaluation criteria}
	\label{Evaluation criteria SS}

Chosen criteria assess the benefits of our method. So, a major point to be evaluated is the correspondence between estimated pure material reflectance spectra and spectra really present in each pixel. To this end we computed, in each pixel $p$, the spectral angles between these two sets of spectra \cite{dennison_comparison_2004}. The resulting criterion is defined as:
\begin{equation*}
	SAM(p) =  \frac{1}{M} \sum_{m=1}^{M} ( \cos^{-1} \left(\frac{\langle \mathbf{r_{m}}(p), \mathbf{\hat{r}_{m}}(p) \rangle}{\Vert \mathbf{r_{m}}(p) \Vert_{2} \cdot \Vert \mathbf{\hat{r}_{m}}(p) \Vert_{2} } \right) )
\end{equation*}
where $\langle \cdot, \cdot \rangle$ stands for the scalar product, $\mathbf{r_{m}}(p)$ the spectrum of the $m^{th}$ source really present in the $p^{th}$ pixel and $\mathbf{\hat{r}_{m}}(p)$ the estimated one. The spectral angle error $SAM(p)$ is then averaged over all pixels to obtain the mean spectral error, $SAM$.
 
Two other criteria were also computed. The first one is the reconstruction error, $RE$, to evaluate the benefit of our method on the global reconstruction of the image. Like for the spectral error, the reconstruction error is computed at pixel level or averaged over all the image:
\begin{equation*}
RE(p) = \frac{1}{L} \cdot \Vert \mathbf{x_{p}} - \mathbf{\hat{c}_{p}}^{T}\mathbf{\hat{R}}(p) \Vert_{F}.
\end{equation*}
The second one is the mean square error computed on the coefficients, $CE$. As for $RE$ and $SAM$ it can be computed at pixel level or averaged on all image: 
\begin{equation*}
CE(p) = \frac{1}{M} \cdot \Vert \mathbf{c_{p}} - \mathbf{\hat{c}_{p}} \Vert_{F}.
\end{equation*}
Computing these errors is possible for semi-synthetic data for which the sources and mixing parameters are known. 

	\subsection{Results}
	\label{Results SS}
	
\begin{figure}[hbtp!]
	\centering
	\includegraphics[scale=0.5]{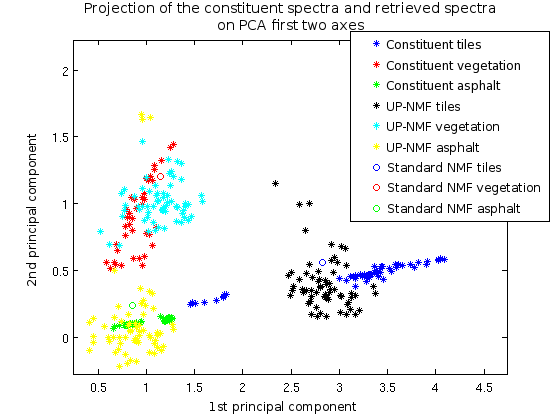}
	\caption{Projection onto the first two PCA axes of constituent spectra (blue, red, green stars), UP-NMF spectra (black, cyan, yellow stars) and standard NMF spectra (blue, red, green circles).}	
	\label{Results fig.UPNMF}
\end{figure}

\begin{figure}[hbtp!]
	\centering
	\includegraphics[scale=0.5]{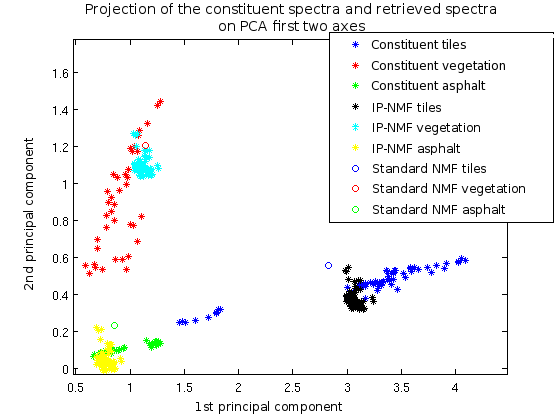}
	\caption{Projection onto the first two PCA axes of constituent spectra (blue, red, green stars), IP-NMF spectra (black, cyan, yellow stars) with $\mu = 30$ and standard NMF spectra (blue, red, green circles).}	
	\label{Results fig.IPNMF 30}
\end{figure}

\begin{figure}[hbtp!]
	\centering
	\includegraphics[scale=0.5]{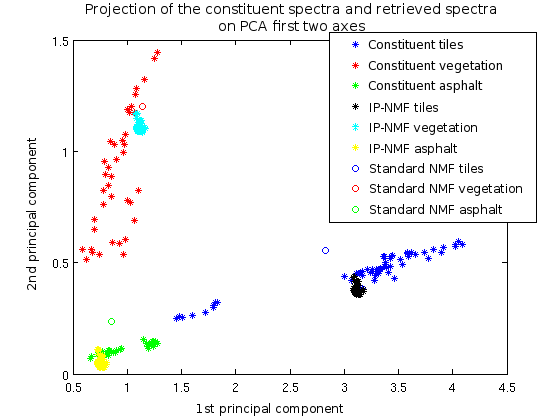}
	\caption{Projection onto the first two PCA axes of constituent spectra (blue, red, green stars), IP-NMF spectra (black, cyan, yellow stars) with $\mu = 100$ and standard NMF spectra (blue, red, green circles).}	
	\label{Results fig.IPNMF 100}
\end{figure}

Figure~\ref{Results fig.UPNMF} illustrates the sources used in the mixing (blue, red and green stars) compared with UP-NMF results (black, cyan and yellow stars) and standard NMF results (blue, red and green circles). Figure~\ref{Results fig.IPNMF 30} and Figure~\ref{Results fig.IPNMF 100} illustrate the same comparisons for IP-NMF with $\mu$ equal to $30$ and $100$ respectively. For these three figures,  $\mathbf{\tilde{R}}^{(0)}$ is built with N-FINDR results and $\mathbf{\tilde{C}}^{(0)}$ with constant coefficients $\frac{1}{M}$.

For each class, the scatter plots of the actual and extracted spectra should be superimposed up to scale factors. 
Indeed scale factors can be contained in the estimated spectra and respectively the inverse in the associated estimated abundances:
\begin{equation}
	\label{Unmixing Methods.NMF pixel by pixel eq1}
	\mathbf{x_{p}} = \sum_{m = 1}^{M} \frac{1}{k_m(p)} c_{pm} \ \times \ k_m(p)\mathbf{r_{m}}(p) \quad \forall p \in \lbrace 1, \ldots, P \rbrace.
\end{equation}
And hence dots can move onto the axes passing through the origin. Thus the dot position onto its class main axis depends on its estimated scale factor.
\begin{figure}[hbtp!]
	\centering
	\includegraphics[scale=0.5]{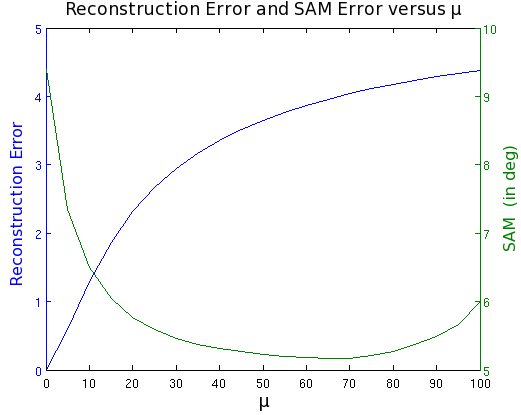}
	\caption{Evolution of the RE (blue curve) and the SAM (green curve) as a function of $\mu$ parameter}	
	\label{Lambda variations}
\end{figure}

As mentioned in Section~\ref{UP-NMF}, the main risk is spectra spreading. The magnitude of the spreading of a class can be measured by the variance of the projection of this class plot onto the line which contains the center of gravity of that class plot and which is orthogonal to the line which contains that center of gravity and the origin. 
This spreading clearly appears in Figure~\ref{Results fig.UPNMF}, the above defined variances for the retrieved spectra are high as compared with the variances for the spectra used in the mixing. This reinforces our idea to limit the class inertia. Results in Table~\ref{Tab1} confirm this observation. In this table, results of UP-NMF and IP-NMF (with $\mu$ fixed to $30$ and $100$) were compared to standard NMF and a classical unmixing method (N-FINDR \cite{nfindr_algo} + FCLS \cite{heinz_fully_2001}). The average SAM of the UP-NMF method is the highest due to the poor reconstruction of some spectra. However we can argue that the Reconstruction Error ($RE$) is the best for this method. But, as for NMF, the $RE$ of UP-NMF have to be carefully interpreted. Indeed $RE$ minimisation is what UP-NMF and standard NMF look for, as shown in (\ref{J_NMF}) and (\ref{J_UPNMF}). So we can obtain good $RE$ and poor $SAM$ and $CE$ with these methods. 
In Figure~\ref{Results fig.IPNMF 30} and Figure~\ref{Results fig.IPNMF 100}, the evolution of the scatter plot with respect to $\mu$ shows the impact of the inertia constraint. If $\mu$ is very high, the IP-NMF method gets closer to a standard unmixing method since it yields very compact sets of spectra for each class (close to a single spectrum). In the case when $\mu$ is suitable, the widths (related to the above mentioned variance) of the scatter plots formed by the estimated spectra are comparable to those of the scatter plots formed by the constituent spectra. That is why $\mu = 30$ is a suitable parameter value: it yields a relevant variance orthogonally to the main axis of the class. Since scale factors do not affect SAM, computing these SAM allows one to confirm this point. Table~\ref{Tab1} shows improvements brought by IP-NMF in particular concerning the $SAM$. Indeed with IP-NMF an improvement of 2 degrees is observed.
Results are given for only 3 values of $\mu$ ($0$ or UP-NMF, $30$ and $100$). They were selected to describe three examples of visually different results. However, as shown by Figure~\ref{Lambda variations}, the evolutions of the $RE$ and $SAM$ are smooth with respect to $\mu$. That means that $\mu$ is not a highly sensitive parameter. It would be possible to fix it to $50$ and obtain similar results. Even $\mu$ equal to $100$ gives better results than the standard NMF and the classical method according to Table \ref{Tab1}. The computational times of UP-NMF and IP-NMF are significantly higher than those of these two standard methods. However the structure of the proposed algorithm makes it quite easily parallelisable. The time variations between the 3 cases are caused by the fluctuations of the number of iterations.\\
%
% \ref{Results fig. 1}
%
\begin{table}[hbtp!]
\caption{Global comparison of unmixing methods.}
\begin{tabular}{@{}cccccc@{}}
\toprule
\multicolumn{1}{l}{}                                            & \begin{tabular}[c]{@{}c@{}}N-FINDR \\ +FCLS\end{tabular} & \begin{tabular}[c]{@{}c@{}}Standard\\ NMF\end{tabular} & UP-NMF & \begin{tabular}[c]{@{}c@{}}IP-NMF \\ ($\mu = 30$)\end{tabular} &  \begin{tabular}[c]{@{}c@{}}IP-NMF \\ ($\mu = 100$)\end{tabular} \\ \midrule \vspace{0.1cm}
$RE$& $17.5$ & $0.6$ & $< 0.1$ & $2.9$ & $4.4$ \\ \vspace{0.1cm}
$SAM$ (deg) & $7.7$ & $7.7$ & $9.4$ & $5.5$ & $6.1$ \\ \vspace{0.1cm}
$CE$ (\%) & $4.0$ & $4.7$ & $3.8$ & $3.8$ & $3.8$\\ \vspace{0.1cm}
Time (sec) & $0.03$ & $0.02$ & $5.87$ & $5.90$ & $6.48$\\
\bottomrule
\end{tabular}
\vspace{0.2cm}
\label{Tab1}
\end{table}

The second point consists in analysing the impact of initialisation. We applied IP-NMF with the initialisation scenarios described in Sec.~\ref{Test description SS}. It appears that a poor initialisation (scenario (i)) leads to poor results for both the standard NMF, UP-NMF and IP-NMF. Yet, compared with standard NMF, IP-NMF improves the average spectral angle error in every initialisation case. We also noted that if both the spectra and the abundance coefficients are initialised too close to a local minimum (scenario (iii) and (b)), the results of our methods are close to this initialisation.

%% file: Test_results_real_image.tex
% Test Results on real images
%
%
	\subsection{Data set}
	\label{Data set}

\begin{figure}[hbtp!]
	\centering
	\includegraphics[scale=0.25]{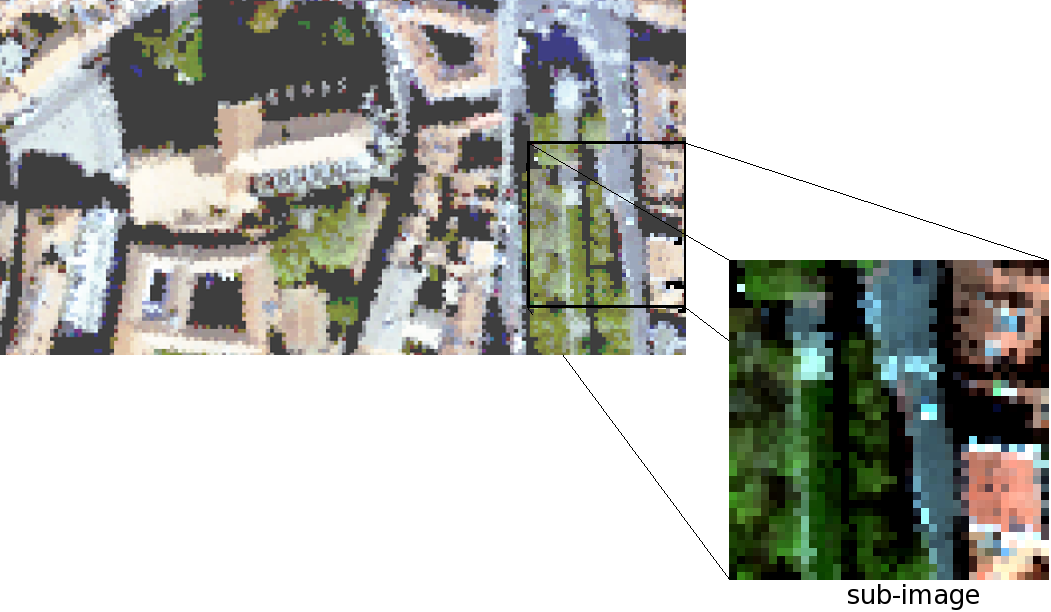}
	\caption{Extracted sub-image.}	
	\label{Sous-images St Etienne }
\end{figure}

In this study we focus on city center image. A sub-image was extracted from the larger image depicted in Figure~\ref{Image St Etienne} and described in Sec.~\ref{Data description}. It contains an avenue, vegetation (trees and grass), tile roofs and shadows of these materials. Figure~\ref{Sous-images St Etienne } depicts this sub-image.

\vspace{0.5cm}
\begin{figure}[hbtp!]
	\centering
	\includegraphics[scale=0.15]{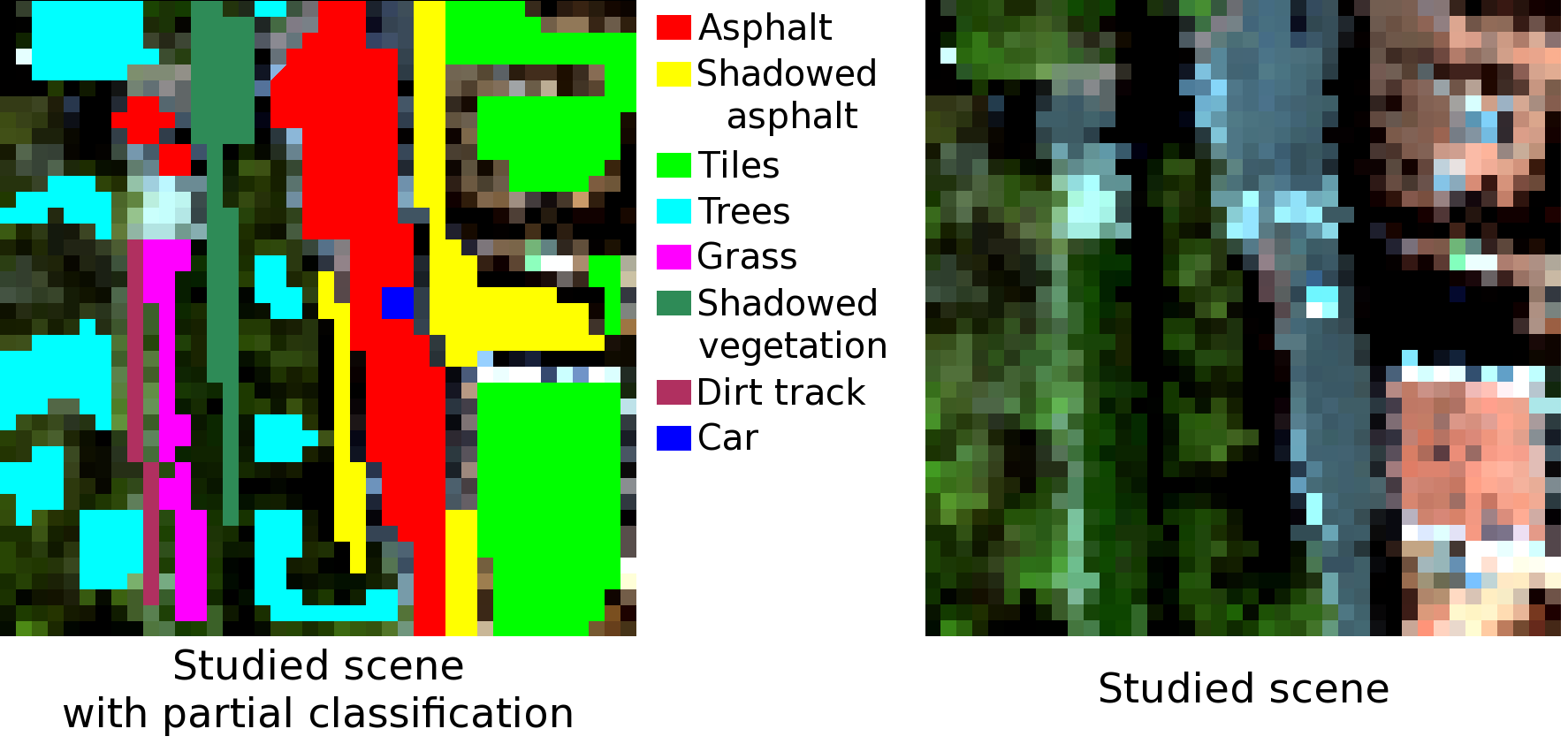}
	\caption{A zoom of the studied scene (right hand side) in real colors ([$647$nm $542$nm $454$nm]) and a manual partial classification of this scene (left hand side). The manual partial classification allows a better understanding of the selected area.}	
	\label{Classif manuelle}
\end{figure}

%\newpage	
	\subsection{Test description}
	\label{Test description DR}
	
Several methods were tested on this data set: 
\begin{itemize}
	\item[$\bullet$] Two classical geometric methods: VCA \cite{nascimento_vertex_2005} and N-FINDR \cite{winter_n-findr:_1999} extract endmembers (one per class of materials). The abundances are then retrieved by a usual Fully Constrained Least Square (FCLS) algorithm \cite{zhang_hybrid_2010}, \cite{broadwater_hybrid_2007}. 
	\item[$\bullet$] A standard NMF method \cite{lin_projected_2007} is applied to retrieve at the same time both the source spectra (one per class) and their associated abundance coefficients.
	\item[$\bullet$] IP-NMF is applied to obtain one set of endmembers per pixel and the associated coefficients.  
\end{itemize} 	
IP-NMF asks for the number of classes, $M$. To find it a well-known method, HySime, developed by Bioucas-Dias et al. in \cite{bioucas-dias_hyperspectral_2008}, was tested to identify the number of endmembers in the hyperspectral image. The results of HySime lead to 46 endmembers. This over-determination of the endmember extraction is due to the intra-class variability. Indeed HySime does not take this phenomenon into account (at least does not enough). The aim of IP-NMF is to work with a limited and realistic number of classes. So this kind of algorithm cannot be used to initialise $M$. It will be manually initialised. In these tests IP-NMF is applied with various parameter initialisations. 
Figure \ref{Classif manuelle} shows that the observed scene mainly consists of three classes: asphalt (red and yellow classes in Fig.~\ref{Classif manuelle}), vegetation (cyan, magenta and dark green in Fig.~\ref{Classif manuelle}) and tiles (green class in Fig.~\ref{Classif manuelle}). Some particular elements could be problematic (the car (blue class in Fig.~\ref{Classif manuelle}), the track (purple class in Fig.~\ref{Classif manuelle}), the roof elements (mixed in the tile pixels)...)
%It seems that the observed scene is composed of asphalt (road), vegetation (trees and grass on the left side of the sub-image of Fig.~\ref{Sous-images St Etienne }) and tiles (roofs). 
So the number of endmembers, $M$, is successively fixed to 3, 5 and 7. In order to perform a fully automated algorithm the initialisation of spectra is firstly obtained by VCA. As a comparison, a second initialisation with manually selected spectra is also performed. Thanks to this manual selection of pure spectra we obtain one of the best possible initialisations. The coefficients are initialised with the constant value $\frac{1}{M}$ since this initialisation was the one yielding the best results for the semi-synthetic data tests (cf.~Subsection.~\ref{Results SS}). 
Considering Fig.~\ref{Lambda variations}, the $SAM$ only moderately varies when $\mu$ ranges between 20 and 80. So a constant $\mu$ was fixed for all the following tests, equal to $30$, as for the semi-synthetic data tests.    

Evaluation criteria cannot be the same as those used for the semi-synthetic data test. Indeed due to the lack of ground truth we cannot compute the $SAM$ and $CE$ errors. Therefore only the Reconstruction error, $RE$ and a usual analysis of the results in various ways (PCA projection, abundance maps...) can be considered. 
It was decided not to exploit the $RE$ because of the reasons developed in Section~\ref{Results SS}.
  
	\subsection{Results}
	\label{Results DR}

\begin{figure}[ht]
	\centering
	\includegraphics[scale=0.60]{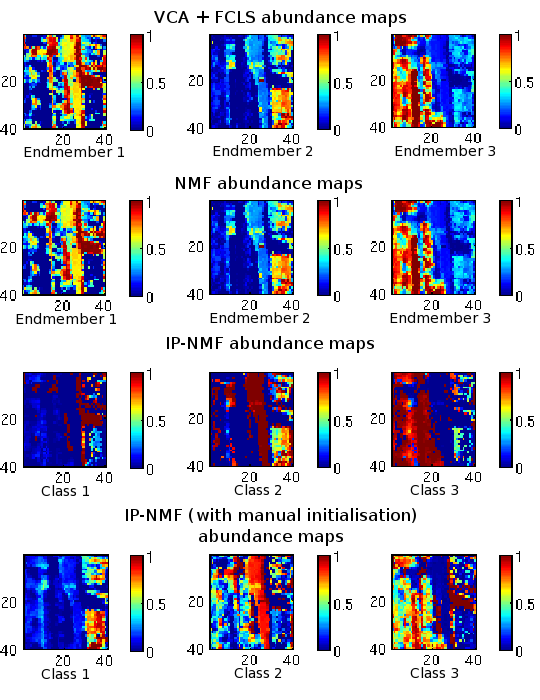}
	\caption{Abundance maps obtained for $M = 3$ with 4 methods: VCA + FCLS ($1^{st}$ line), standard NMF ($2^{nd}$ line), IP-NMF initialised with VCA ($3^{rd}$ line) and IP-NMF initialised with manually selected spectra ($4^{th}$ line).}	
	\label{Comparaison resultats 2m 3classes}
	%\vspace{-0.5cm}
\end{figure}

\begin{figure*}[t]
	\centering
	\includegraphics[scale=0.5]{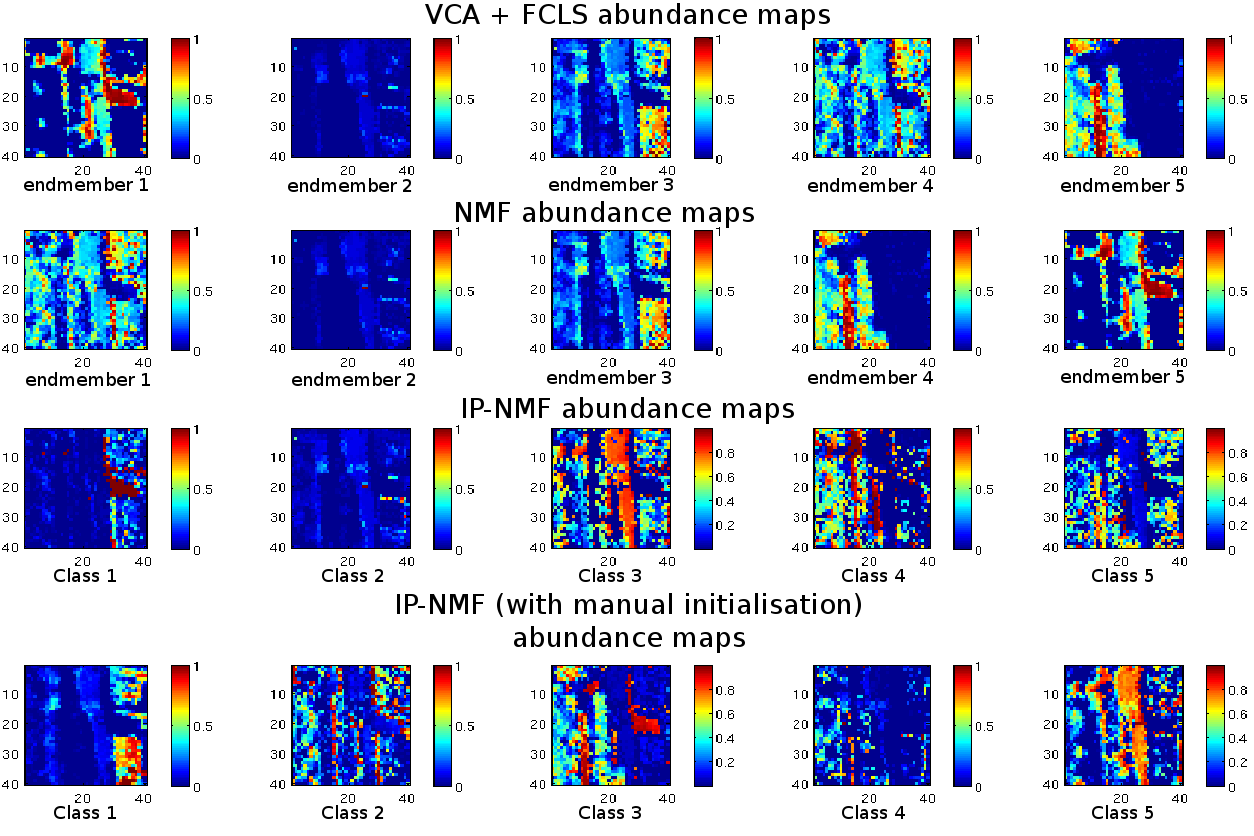}
	\caption{Abundance maps obtained for $M = 5$ with 4 methods: VCA + FCLS ($1^{st}$ line), standard NMF ($2^{nd}$ line), IP-NMF initialised with VCA ($3^{rd}$ line) and IP-NMF initialised with manually selected spectra ($4^{th}$ line).}	
	\label{Comparaison resultats 2m 5classes}
\end{figure*}

\begin{figure*}
	\centering
	\includegraphics[scale=0.30]{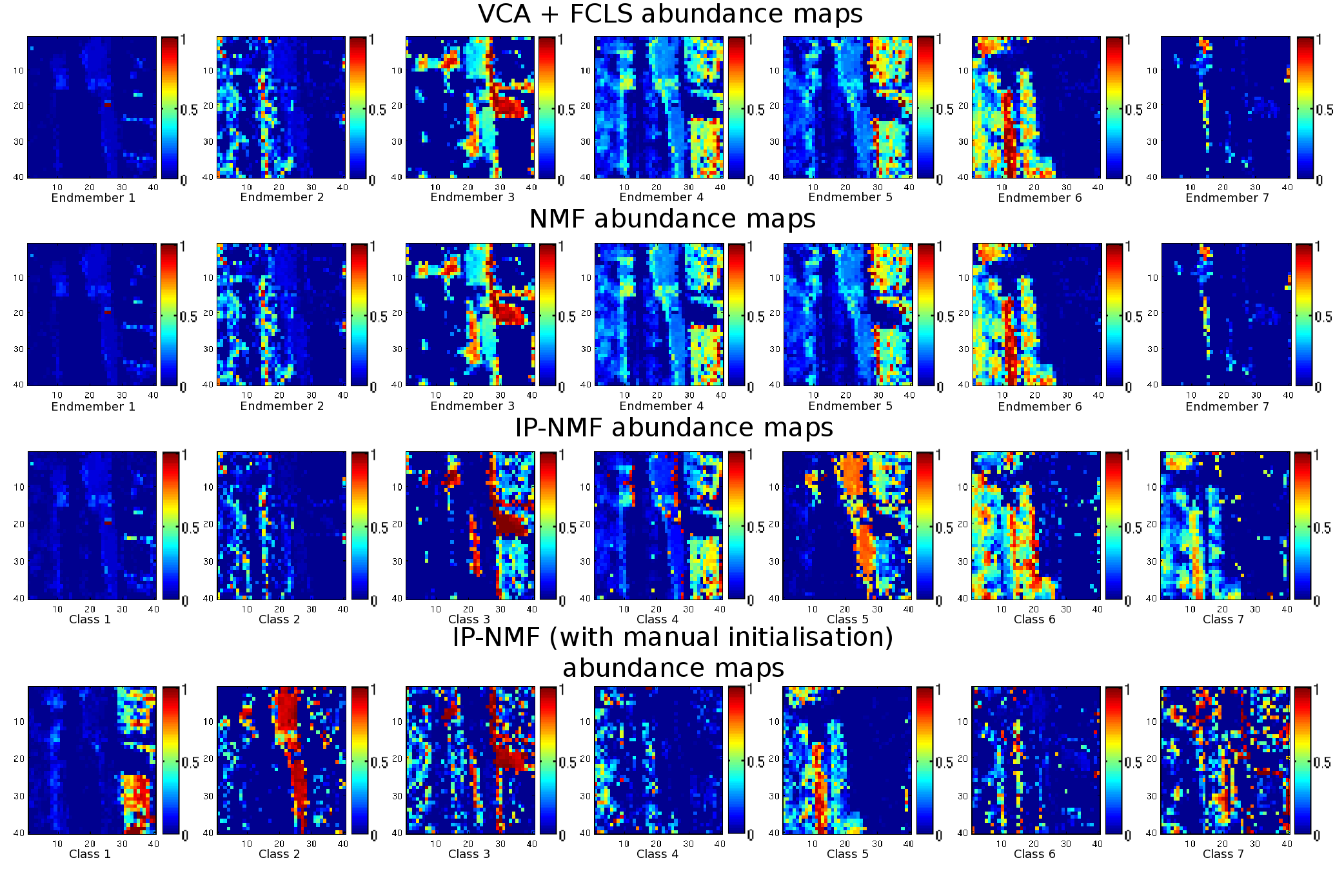}
	\caption{Abundance maps obtained for $M = 7$ with 4 methods: VCA + FCLS ($1^{st}$ line), standard NMF ($2^{nd}$ line), IP-NMF initialised with VCA ($3^{rd}$ line) and IP-NMF initialised with manually selected spectra ($4^{th}$ line).}	
	\label{Comparaison resultats 2m 7classes}
\end{figure*}

IP-NMF was applied with the two initialisations described in the previous subsection. Results  are depicted in Figures~\ref{Comparaison resultats 2m 3classes} to~\ref{Comparaison resultats 2m 7classes}. Each of these figures shows the abundance maps of VCA + FCLS unmixing ($1^{st}$ line), NMF ($2^{nd}$ line), IP-NMF initialised with VCA ($3^{rd}$ line) and IP-NMF manually initialised ($4^{th}$ line) respectively applied with $M$ equal to 3, 5 and 7. The results obtained with N-FINDR are not shown because they are very close to the ones given by VCA.\\
	
%IP-NMF was firstly performed on image with the following initialisation~: (i) 3 classes, (ii-a) spectra initialised with a VCA algorithm or (ii-b) spectra initialised with 3 spectra manually selected in the image, (iii) abundances initialised with constant coefficients. It seems that the observed scene is composed of asphalt (street), vegetation (trees and grass on the image left side) and tiles (roofs) which justify (i). We choose (ii-a) to obtain a fully automatised algorithm, (ii-b) was selected to compare our result with one of the best spectra initialisation. (iii) was chosen because it is the initialisation with the best result in the semi-synthetic tests. VCA and the standard NMF were also performed on this image with the same initialisation condition as IP-NMF. Due to the  Figure~\ref{Comparaison resultats 2m 3classes} shows the resulting abundance maps. 
%%
%
%\vspace{-0.5cm}
% 

			\subsubsection{Results of classical unmixing methods (one set of endmembers per image)}

As described previously, two methods were applied to the image, VCA + FCLS and a standard NMF. Abundance maps are depicted in the $1^{st}$ and $2^{nd}$ lines of Figures~\ref{Comparaison resultats 2m 3classes} to~\ref{Comparaison resultats 2m 7classes}. 
It appears that VCA + FCLS and NMF maps are relatively similar, this is due to the initialisation. Indeed NMF was initialised by VCA resulting spectra. Several observations are made from these maps. 

Firstly the asphalt class (road) is never accurately extracted. In the three figures it is a mixture between shadow endmember and another one. For instance in Figure~\ref{Comparaison resultats 2m 3classes} it is a mixture between tile endmember and shadow endmember. 
Then, increasing the number of endmembers reduces the unmixing quality because these unmixing methods then sometimes focus on specific spectra (vertices of the simplex), instead of extracting all major classes. It is the case in Figure~\ref{Comparaison resultats 2m 5classes} and Figure~\ref{Comparaison resultats 2m 7classes} where a car spectrum is extracted ($2^{nd}$ column and $1^{st}$ column respectively in Fig.~\ref{Comparaison resultats 2m 5classes} and Fig.~\ref{Comparaison resultats 2m 7classes}). Conversely due to intra-class variability several endmembers can be extracted belonging to the same class, for instance the tiles in Figure~\ref{Comparaison resultats 2m 7classes} ($5^{th}$ and $6^{th}$ columns). 

For this image it seems that Figure~\ref{Comparaison resultats 2m 3classes} depicts the best results. Indeed except the road all classes are depicted. However the vegetation ($3^{rd}$ column) is not uniform. Indeed it is the \emph{shadow} spectrum which was used to rebuild the shadowed tree spectra. 

%Contrary to VCA and NMF asphalt class exists after performing IP-NMF indeed on figure , (b) a tile spectrum is extracted with VCA and NMF (endmember 2) but not with IP-NMF, in this case it is included in the $2^{nd}$ class, (c) area where abundances are high seems more homogeneous in IP-NMF than with other two methods, (d) with IP-NMF a class shaded asphalt exists ($1^{st}$ class) and does not include shaded vegetation spectra (in $3^{rd}$ class). 

    	\subsubsection{Results of IP-NMF method initialised with VCA}
    	
To analyse the ``automated" IP-NMF results the $3^{rd}$ line of Figures~\ref{Comparaison resultats 2m 3classes}, \ref{Comparaison resultats 2m 5classes} and \ref{Comparaison resultats 2m 7classes} is studied. 
First of all it has to be noted that even if abundance maps are presented in the same way for VCA, NMF and IP-NMF, they depict, in fact, two different things. Indeed for VCA and NMF each map is associated with a single endmember whereas for IP-NMF a map is associated with a class. This means that each abundance of IP-NMF abundance maps is associated with a spectrum different from its neighbours. The understanding of the results is highly linked to this preliminary remark. 

Our first observation concerning Figure~\ref{Comparaison resultats 2m 3classes} is the homogeneity of areas: contrary to the VCA and NMF results, there is no cyan background in the maps, which would correspond to a material weakly present in a large number of pixels. This is the result of the introduction of degrees of freedom for modelling intra-class variability in our adaptive variables corresponding to $\mathbf{\tilde{C}}$ and $\mathbf{\tilde{R}}$. The three classes obtained with IP-NMF are: shadowed asphalt ($1^{st}$ column), sunny asphalt + tiles ($2^{nd}$ column), vegetation ($3^{rd}$ column). The $2^{nd}$ class contains spectra of asphalt and tiles. This result is the consequence of two things: the low inter-class variability (cf. Sec.~\ref{Data study}) and the initialisation. Indeed, due to the low inter-class variability, constituent spectra are spectrally close even if they do not belong to the same class. So it is possible that, in a same class, estimated spectra evolve in different ways and form two clusters, i.e. two sub-classes inside one class. This is what happens in this test. This is possible since the distance between the two center of gravity of the sub-classes is similar to the intra-class variability of some classes. Initialisation is also accountable for this phenomenon since all the spectra of a class evolved from a same ``seed spectrum''. So, if the initialisation spectrum (or ``seed spectrum'') is close to the two classes this phenomenon more easily appears.

Since this phenomenon is due to the combination of two classes which are spectrally too close, what happens when the number of classes is increased ? This is depicted in Figures~\ref{Comparaison resultats 2m 5classes} and \ref{Comparaison resultats 2m 7classes}. It can be observed that the results are not as expected. Indeed new classes appear, for instance the car (Fig.~\ref{Comparaison resultats 2m 5classes}, $2^{nd}$ column and Fig.~\ref{Comparaison resultats 2m 7classes}, $1^{st}$ column) but the quality of the unmixing decreases. For instance the road is now considered as a mixture of two spectra. This decrease of the quality is due to the initialisation, as will be shown in Section \ref{IP-NMF manuel}. Indeed the VCA spectra are used to initialise IP-NMF and the poor quality of spectra extracted by VCA influences the results of IP-NMF, in which spectra cannot move enough to make up for the initialisation error. However IP-NMF gives better results than VCA or NMF. Indeed some classes are detected only by IP-NMF. Besides sometimes spectra in a pixel are misclassified by IP-NMF (i.e. they do not belong to the same physical class as other spectra of this class) but anyway they accurately fit the actual pure spectra present in this pixel.\\

		\subsubsection{Results of IP-NMF method with manual initialisation}
		\label{IP-NMF manuel}

The $4^{th}$ line of Figures~\ref{Comparaison resultats 2m 3classes},~\ref{Comparaison resultats 2m 5classes} and~\ref{Comparaison resultats 2m 7classes} shows the impact of the initialisation on IP-NMF results. IP-NMF was initialised respectively by 3, 5 and 7 spectra manually selected in the image. This allows one to choose the initial classes and ensures that initialisation spectra are pure ones. In Figure~\ref{Comparaison resultats 2m 3classes}, selected spectra were a tile spectrum, an asphalt spectrum and a tree spectrum. Abundance maps show that the tiles are very well extracted (column 1 of Fig.~\ref{Comparaison resultats 2m 3classes} compared to the sub-image of Fig.~\ref{Sous-images St Etienne }). Asphalt is quite well extracted (column 2 of Fig.~\ref{Comparaison resultats 2m 3classes}), but the road is not as homogeneous as the road extracted by automated IP-NMF (row 3, column 2 of Fig.~\ref{Comparaison resultats 2m 3classes}). The $3^{rd}$ extracted class (column 3 of Fig.~\ref{Comparaison resultats 2m 3classes}) contains vegetation and shadowed asphalt. As in the previous case (automated IP-NMF) abundance maps show a combination of 2 subclasses in a class. With this manual initialisation, an improvement of the result is expected when the number of classes is increased, because the quality of automated IP-NMF results falls due to the poor initialisation given by VCA. In Figure~\ref{Comparaison resultats 2m 7classes}, seven classes of spectra were extracted: tile, asphalt,  shadowed asphalt, tree, grass, shadowed tree, path (resp. columns 1 to 7 of Fig.~\ref{Comparaison resultats 2m 7classes}). Compared to the automated IP-NMF results, the extraction of some classes is better performed here. Tiles are well retrieved despite their high variability, as the asphalt. However the vegetation extraction is better with the automated IP-NMF, indeed the grass ($5^{th}$ column) and tree ($4^{th}$  column) classes compete, which leads to poor tree extraction here. Besides the shadowed trees and shadowed tiles are extracted in the same class ($7^{th}$ column) due to their spectral proximity.

So, it seems that a manual initialisation of the spectra with spectra belonging to expected classes improves the extraction of classes with high intra-class variability (tiles). However when intra-class variability is smaller, this initialisation can lead to lower performance than the automated one. For instance shadowed trees have a small inertia compared with the tiles, so the shadowed tree class can grow and other shadowed spectra are absorbed. However these spectra, even if they belong to the same class, still are spectrally different. \\

		\subsubsection{Result of automated IP-NMF with a post-processing}

\begin{figure*}
	\centering
	\includegraphics[scale=0.30]{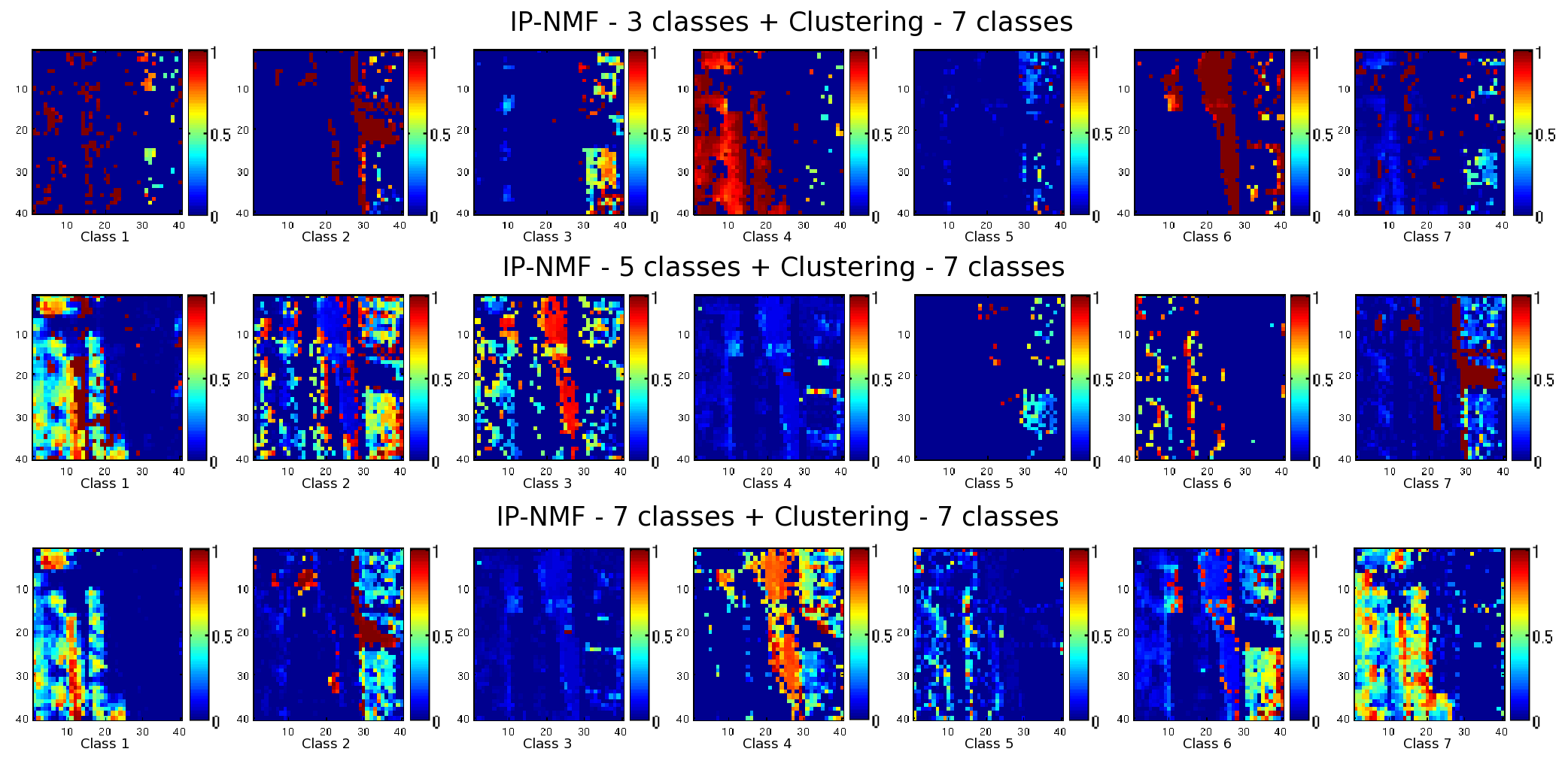}
	\caption{Abundance maps obtained with automated IP-NMF applied with $M = 3$ ($1^{st}$ line), $M = 5$ ($2^{nd}$ line) and $M = 7$ ($3^{rd}$ line) and followed by a clustering with $7$ classes.}	
	\label{Comparaison resultats clustering}
\end{figure*}
		
Even with a good initialisation some problems persist. We propose to solve them by keeping the above automated IP-NMF method, followed by a post-processing stage. We exploit the fact that each pixel is described by a unique set of endmembers. After applying IP-NMF, we gathered all these spectra and then clustered them with the k-means method \cite{macqueen_methods_1967} using spectral angle as the similarity measure. The resulting abundance maps are depicted in Figure~\ref{Comparaison resultats clustering}.

This post processing appears to be all the more efficient as the number of endmembers searched by IP-NMF is small. Indeed modifications of the maps between Fig.~\ref{Comparaison resultats 2m 7classes}, $3^{rd}$ line and Fig.~\ref{Comparaison resultats clustering}, $3^{rd}$ line (with classes extracted in arbitrary orders) are small compared to the differences between Fig.~\ref{Comparaison resultats 2m 3classes}, $3^{rd}$ line and Fig.~\ref{Comparaison resultats clustering}, $1^{st}$ line. This post-processing allows us splitting the classes composed of two subclasses. For instance, in the $M = 3$ case, tile class and asphalt class are well separated. This post processing therefore allows one to further exploit the large amount of information extracted by IP-NMF, by improving the classification of the extracted pure spectra. Other post-processing methods can be imagined for other applications which would exploit the high number of spectra extracted with IP-NMF.